\pdfoutput=1

\documentclass[a4paper,12pt,bold]{scrartcl}

\usepackage{color,colortbl}																			%Farbige Tabellen
\usepackage{xcolor}

\usepackage{apacite}
\usepackage{footnote}
\makesavenoteenv{tabular}
\usepackage{makecell}

\usepackage{enumitem}

\usepackage{ marvosym}
\usepackage{threeparttablex}

\usepackage[makeroom]{cancel}
\normalsize

\newcommand{\btheta}{\bm{\theta}}
\newcommand{\bTheta}{\bm{\Theta}}
\newcommand{\G}{\mathcal{G}}	%Normal Distribution

\newcommand*\diff{\mathop{}\!\mathrm{d}}
\newcommand{\N}{\mathcal{N}}	%Normal Distribution
\newcommand{\U}{\mathcal{U}}	%Uniform Distribution
\newcommand{\E}{\mathrm{E}}		%Expectation
\newcommand{\bs}{\boldsymbol}

\usepackage{bbm}

\usepackage{algpseudocode,tabularx,ragged2e}
\newcolumntype{C}{>{\centering\arraybackslash}X} % centered "X" column
\newcolumntype{L}{>{\arraybackslash}X} % centered "X" column

\usepackage{algorithmicx}

\usepackage{algorithm}

\let\Algorithm\algorithm
\renewcommand\algorithm[1][]{\Algorithm[#1]\setstretch{1.5}}

\newcommand{\M}{\mathcal{M}}

\definecolor{lightgrey}{gray}{0.90}	%Farben mischen
\definecolor{grey}{gray}{0.85}
\definecolor{darkgrey}{gray}{0.65}
\definecolor{lightblue}{rgb}{0.8,0.85,1}

\newcolumntype{g}{>{\columncolor{gray}}c}
\usepackage{bibentry}
\usepackage{booktabs}
\usepackage{epigraph}
\usepackage[sans]{dsfont}

\usepackage{bm}																									%matrix symbol
\usepackage{setspace}																						%Fu�noten (allgm.
\usepackage[colorlinks = true,
            linkcolor = blue,
            urlcolor  = blue,
            citecolor = blue,
            anchorcolor = blue]{hyperref}%Zeilenabst�nde)
\usepackage{threeparttable}
\usepackage{lscape}																							%Querformat
\usepackage[latin1]{inputenc}																		%Umlaute
\usepackage{graphicx}
\graphicspath{{material/}{material/structural-model/}}

\makeatletter
\def\@mb@citenamelist{cite,citep,citet,citealp,citealt,citepalias,citetalias}
\makeatother

\usepackage[round,longnamesfirst]{natbib}
\usepackage[resetlabels]{multibib}

\newcites{Appndx}{References Appendix}

\makeatletter
\providecommand\@newciteauxhandle{\@auxout}
\def\@restore@auxhandle{\gdef\@newciteauxhandle{\@auxout}}
\AtBeginDocument{%
\@ifundefined{newcites}{\global\let\@restore@auxhandle\relax}{}}
\makeatother

\usepackage{amsmath, placeins}
\usepackage{amssymb}
\usepackage{fancybox}																						%Boxen und Rahmen
\usepackage{appendix}
\usepackage{listings}
\usepackage{xr}

\usepackage{enumerate}

\usepackage{tabularx}
\usepackage{longtable,tabu}
\usepackage{subfig,float}																				%Mehrseitige Tabellen
\usepackage{color,colortbl}																			%Farbige Tabellen
\usepackage[left=2.0cm, right=2.0cm, top=2.0cm, bottom=2.0cm,foot=1.cm]{geometry} %Seitenr�nder
\definecolor{lightgrey}{gray}{0.95}	%Farben mischen
\definecolor{grey}{gray}{0.85}
\definecolor{darkgrey}{gray}{0.80}

\usepackage{tikz}
\usetikzlibrary{positioning}

\usepackage[labelfont=bf]{caption}
\usepackage{url}  % Used for linebreaks in verbatim statements

\newtheorem{Remark}{Remark}

\deffootnote[1em]{1.1em}{0em}{\textsuperscript{\thefootnotemark}}

\DeclareMathOperator*{\argmin}{arg\,min}
\DeclareMathOperator*{\argmax}{arg\,max}



\newcommand{\ind}{{\textbf{I}}}

\definecolor{tabblue}{RGB}{31,119,180}  % usage: blue-collar occupation
\definecolor{taborange}{RGB}{255, 127, 14}  % usage: school
\definecolor{tabgreen}{RGB}{44, 160, 44}  % usage: home
\definecolor{tabred}{RGB}{214, 39, 40}  % usage: white-collar occupation
\definecolor{tabpurple}{RGB}{148, 103, 189}  % usage: military
\definecolor{tabbrown}{RGB}{140, 86, 75}
\definecolor{tabrose}{RGB}{227, 119, 194}
\definecolor{tabgrey}{RGB}{127, 127, 127}
\definecolor{tablime}{RGB}{188, 189, 34}
\definecolor{tabcyan}{RGB}{23, 190, 207}

\definecolor{bwtabblue}{RGB}{77, 77, 77}
\definecolor{bwtabred}{RGB}{42, 42, 42}
\definecolor{bwtabpurple}{RGB}{130, 130, 130}
\definecolor{bwtaborange}{RGB}{209, 209, 209}
\definecolor{bwtabgreen}{RGB}{187 ,187 ,187}

\title{Structural models for policy-making\thanks{Corresponding author: Philipp Eisenhauer, peisenha@uni-bonn.de. Philipp Eisenhauer and Lena Janys are both funded by the Deutsche Forschungsgemeinschaft (DFG, German Research Foundation) under Germany's Excellence Strategy - EXC 2126/1- 390838866 and the TRA Modelling (University of Bonn) as part of the Excellence Strategy of the federal and state governments. Jano\'s Gabler is grateful for financial support by the German Research Foundation (DFG) through CRC-TR 224 (Project C01) and funding by IZA Institute of Labor Economics. Lena Janys is funded by the Deutsche Forschungsgemeinschaft (DFG, German Research Foundation) under Germany's Excellence Strategy - EXC 2047/1 - 390685813. Philipp Eisenhauer was funded by a postdoctoral fellowship by the AXA Research Fund. We thank Tim Mensinger for his help in the early stages of the project. We thank Max Blesch, Joachim Freyberger, Annica Gehlen, Daniel Harenberg, Ken Judd, Gregor Reich, J\"org Stoye, and Rafael Suchy for numerous helpful discussions. We thank Michael Keane and Kenneth Wolpin for providing the dataset used in our analysis. We thank Annica Gehlen and Emily Schwab for their outstanding research assistance. We are grateful to the Social Sciences Computing Service (SSCS) at the University of Chicago for the permission to use their computational resources. Eisenhauer: University of Bonn, peisenha@uni-bonn.de; Gabler: University of Bonn and IZA, janos.gabler@uni-bonn.de, Janys: University of Bonn, ljanys@uni-bonn.de}}
\subtitle{Coping with parametric uncertainty}
\author{Philipp Eisenhauer \& Jano\'s Gabler \& Lena Janys \& Christopher Walsh\\ University of Bonn\vspace{0.5cm}}
\date{\today}

\graphicspath{{material/}}

\begin{document}

% !TEX root = ../main.tex

\maketitle

\setcounter{page}{1}
\thispagestyle{empty}
\begin{abstract}\noindent
The ex-ante evaluation of policies using structural econometric models is based on estimated parameters as a stand-in for the true parameters. This practice ignores uncertainty in the counterfactual policy predictions of the model. We develop a generic approach that deals with parametric uncertainty using uncertainty sets and frames model-informed policy-making as a decision problem under uncertainty. The seminal human capital investment model by \citet{Keane.1997} provides a well-known, influential, and empirically-grounded test case. We document considerable uncertainty in the models's policy predictions and highlight the resulting policy recommendations obtained from using different formal rules of decision-making under uncertainty.\\
\end{abstract}

\noindent\begin{tabular}{@{\hspace{0.5cm}}ll}
\textbf{JEL Codes} & D81, C44, J01\\
\textbf{Keywords}  & decision-making under uncertainty, structural microeconometrics
\end{tabular}

\newpage\tableofcontents\clearpage

%!TEX root = ../main.tex

\setcounter{page}{1}
%---------------------------------------------------------------------------------------------------
\FloatBarrier\section{Introduction}
%---------------------------------------------------------------------------------------------------
Structural microeconometricians use highly parameterized computational models to investigate economic mechanisms, predict the impact of proposed policies, and inform optimal policy-making \citep{Wolpin.2013}. These models represent deep structural relationships of theoretical economic models invariant to policy changes \citep{Hood.1953}. The sources of uncertainty in such an analysis are ubiquitous \citep{Saltelli.2020}. For example, models are often misspecified, there are numerical approximation errors in their implementation, and model parameters are uncertain. Therefore, most disciplines require a proper account of uncertainty before using computational models to inform decision-making \citep{Council.2012,SAPEA.2019}.\\

\noindent The following study focuses on parametric uncertainty in structural microeconometric models that are estimated on observed data. Researchers often do not account for parametric uncertainty and conduct an as-if analysis in which the point estimates serve as a stand-in for the true model parameters. They then continue to study the implications of their models at the point estimates \citep{Adda.2017,Blundell.2016,Eckstein.2019,Eisenhauer.2015b} and rank competing policy proposals based on the point predictions alone \citep{Blundell.2012,Cunha.2010,Gayle.2019,Todd.2006}. In fact, \citet{Keane.2011d} states in their handbook article that they are unaware of any applied work that reports the distribution of policy predictions under parametric uncertainty. To the best of our knowledge, this statement remains true more than a decade later. Consequently, economists risk accepting fragile findings as facts, ignoring the trade-off between model complexity and prediction uncertainty, and neglecting to frame policy advice as a decision problem under uncertainty.\\

\noindent To mitigate these shortcomings, we develop an approach that copes with parametric uncertainty in structural microeconometric models and embeds model-informed policy-making in a decision-theoretic framework. Ideally, policy-makers fix the parameter space ex-ante and then evaluate the policy options according to decision rules. However, this approach is often computationally intractable. We, therefore, follow \citet{Manski.2021}'s suggestion and, instead of using the parameter estimates as-if they were true, incorporate uncertainty in the analysis by treating the estimated confidence set as-if it is correct. We use the confidence set to construct an uncertainty set that is anchored in empirical estimates, statistically meaningful, and computationally tractable \citep{Ben-Tal.2013}. Instead of just focusing on the point estimates, we evaluate counterfactual policies based on all parametrizations within the uncertainty set.\\

\noindent We draw on statistical decision theory \citep{Manski.2013} to deal with the uncertainty in counterfactual predictions. This approach promotes a well-reasoned and transparent policy process. Before a decision, it clarifies trade-offs between choices \citep{Gilboa.2018}. Afterward, decision-theoretic principles allow constituents to scrutinize the coherence of choices \citep{Gilboa.2020}, ease the ex-post justification \citep{Berger.2021}, and facilitate the communication of uncertainty \citep{Manski.2019}.\\

\noindent We tailor our approach to the class of Eckstein-Keane-Wolpin (EKW) models \citep{Aguirregabiria.2010}. Labor economists often use EKW models to learn about human capital investment and consumption-saving decisions and predict the impact of proposed reforms to education policy and welfare programs \citep{Keane.2011d,Low.2017,Blundell.2017}. The analysis of these models poses serious computational challenges. During estimation, EKW models are solved thousands of times and even a single solution often takes several minutes. Thus, a decision-theoretic ex-ante analysis of alternative decision rules across the whole parameter space, as intended by \citep{Wald.1950}, is infeasible. Instead we construct an uncertainty set, a subset of the whole parameter space, and deal with the ex-post uncertainty after estimating the model. This compromise allows us to garner the benefits of using statistical decision theory to shape policy-making under uncertainty while ensuring the computational tractability of our analysis.\\

\noindent As an example of our approach, we analyze the seminal human capital investment model by  \citet{Keane.1997} as a well-known, empirically grounded, and computationally demanding test case. We follow the authors and estimate the model on the National Longitudinal Survey of Youth 1979 (NLSY79) \citep{NLSY.2019} using the original dataset and reproduce all core results. We revisit their predictions for the impact of a tuition subsidy on completed years of schooling. The economics of the model implies that the nonlinear mapping between the model parameters and predictions is truncated at zero, and we thus use the Confidence Set (CS) bootstrap \citep{Woutersen.2019} to estimate the confidence set for the counterfactuals. We document considerable uncertainty in the policy predictions and highlight the resulting policy recommendations from different formal rules on decision-making under uncertainty.\\

\noindent Our work extends existing research exploring the sensitivity of implications and predictions to parametric uncertainty in macroeconomics and climate economics. For example, \citet{Harenberg.2019} study uncertainty propagation and sensitivity analysis for a standard real business cycle model. \citet{Cai.2019} examine how uncertainties and risks in economic and climate systems affect the social cost of carbon. However, neither of them estimates their model on data. Instead, they rely on expert judgments to inform the degree of parametric uncertainty. They do not investigate the consequences of uncertainty for policy decisions in a decision-theoretic framework.\\

\noindent We complement a burgeoning literature on the sensitivity analysis of policy predictions in light of model or moment misspecification. For example, \citet{Andrews.2017} and \citet{Andrews.2020} treat the model specification as given and then analyze the sensitivity of the parameter estimates to the misspecification of the moments used for estimation. \citet{Christensen.2019} study global sensitivity of the model predictions to misspecification of the distribution of unobservables. \citet{Jorgensen.2021} provides a local measure for the sensitivity of counterfactuals to model parameters that are fixed before the estimation of the model.\footnote{For other examples, see \citet{Armstrong.2021}, \citet{Bonhomme.2020}, \citet{Bugni.2019}, and \citet{Mukhin.2018}.} This literature does not embed the counterfactual predictions in a decision-theoretic setting. Recent work by \citet{Kalouptsidi.2020}, \citet{Kalouptsidi.2021}, and \citet{Norets.2014} studies (partial) identification and inference on counterfactuals. However, they all adopt the setup outlined in \citet{Rust.1987} and exploit the additive separability of the immediate utility function between observed and unobserved state variables, which does not apply to EKW models. In related work, \citet{Blesch.2021} conduct a decision-theoretic ex-ante analysis to determine optimal decision rules in \citet{Rust.1987}'s stochastic dynamic investment model where the decision-maker directly accounts for uncertainty in the model's transition dynamics. They only consider uncertainty in a subset of the model's parameters which are estimated outside the model and remain fixed to their point estimates during the analysis.\\

\noindent In Section \ref{Framework}, we describe the decision-theoretic framework for making model-informed decisions under parametric uncertainty using an illustrative example. After summarizing the empirical setting of \citet{Keane.1997} in Section \ref{Setting}, we present our results in Section \ref{Results}.  We complete our analysis in Section \ref{Conclusion} with a brief conclusion and outlook.

%!TEX root = ../main.tex
%---------------------------------------------------------------------------------------------------
\FloatBarrier\section{Structural models for policy-making}\label{Framework}
%---------------------------------------------------------------------------------------------------
In the following section, we discuss uncertainty propagation and the common practice of using estimated parameters as a plug-in replacement for the true model parameters. We then explore the limitations of this strategy and introduce our alternative approach, in which we implement estimated confidence sets to construct uncertainty sets. In so doing, we are able to cope with uncertain policy predictions in a proper decision-theoretic framework. \\

\noindent At a high level, a structural microeconometric model provides a mapping $\M(\btheta)$ between the $l$ model parameters $\btheta \in \bTheta$ and a quantity $y$ that is of interest to policy-makers.
\begin{align*}
  \mathbb{R}^l \supset \bTheta \ni \btheta \mapsto  \M(\btheta) = y
\end{align*}
\noindent A policy $g \in \G$ changes the mapping to $\M_g(\btheta)$ and produces a counterfactual $y_g$.\\

\noindent Estimation of a baseline model $\M(\btheta)$ describing the status-quo on observed data allows researchers to learn about the true parameters. Frequentist estimation procedures such as maximum likelihood estimation and the method of simulated moments produce a point estimate $\hat{\btheta}$. However, uncertainty about the true parameters remains.\\

\noindent Previewing our empirical analysis of \citet{Keane.1997}, our $\M$ is provided by a dynamic model of human capital accumulation, which we estimate on observed schooling and labor market decisions using simulated maximum likelihood estimation. The policy $g$ is the implementation of a college tuition subsidy, and the counterfactual is the level of completed schooling in the population. Example parameters that drive the economics of the model are time preferences of individuals, the return to schooling, and the transferability of work experience across occupations.\\

\noindent The following illustrative example highlights our key points. We consider two policies $g \in\{1, 2\}$  that result in two different mappings $ (\M_1, \M_2)$ of the same scalar $\theta$ to a counterfactual $y_g$. Higher values of $y_g$ are more desirable for a policy-maker. The point estimate $\hat{\theta}$ is determined by estimating a baseline model on an observed dataset. We denote the probability density function of its sampling distribution by $f_{\hat{\theta}}$.\\

\noindent Under the first policy, the counterfactual is an increasing nonlinear function of $\theta$. In the case of the second policy, the relationship is decreasing and linear.
\begin{figure}[ht!]\centering
\scalebox{0.35}{\includegraphics{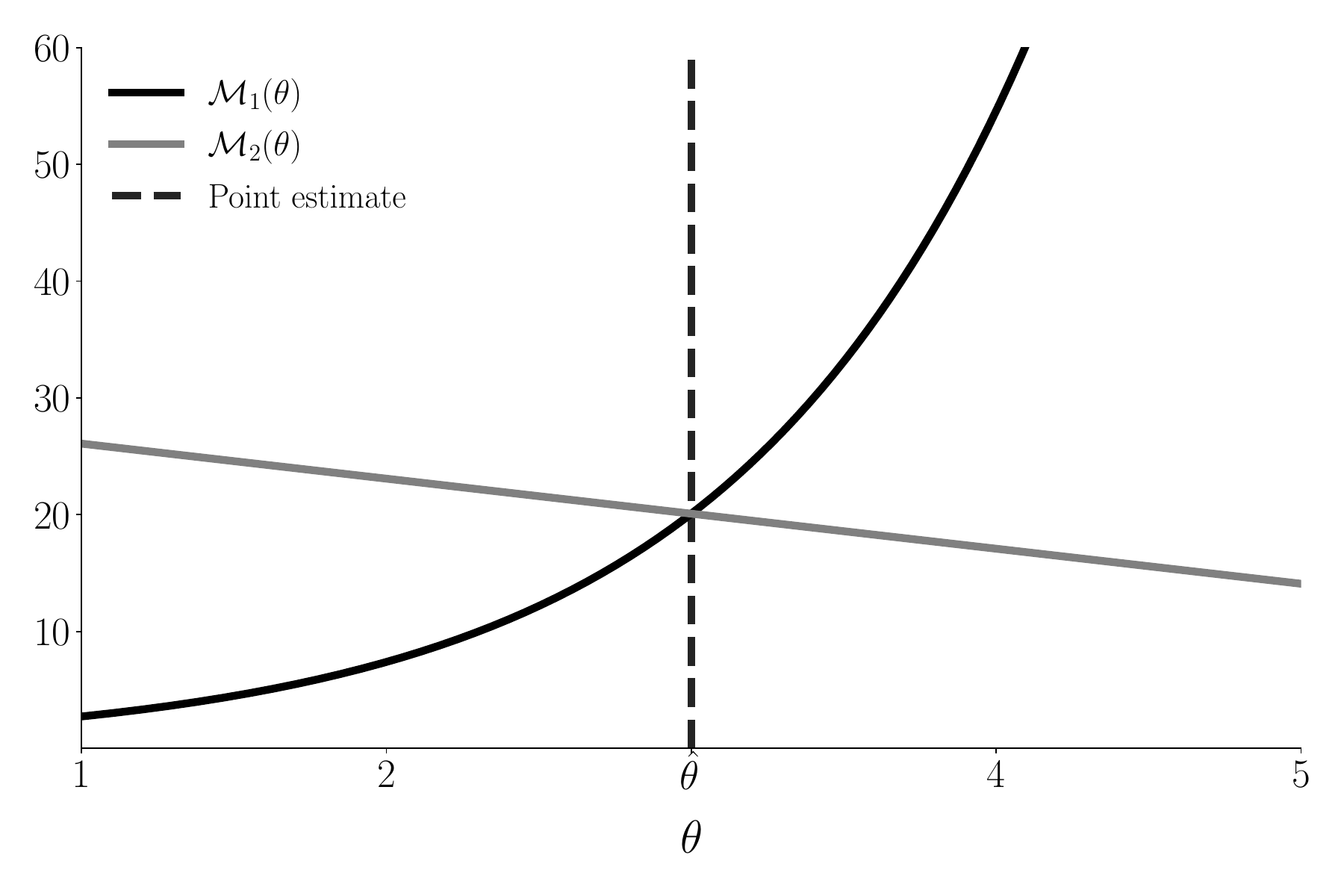}}
\caption{Model comparison}\label{Illustration model comparison}
\begin{center}
\begin{minipage}[t]{0.6\columnwidth}
\item \scriptsize{\textbf{Notes:} We parameterize the two models as $y_1  = \exp{\theta}$ and $y_2 = 29.08 -3\, \theta$.}
\end{minipage}\end{center}
\end{figure}\FloatBarrier
\noindent Figure \ref{Illustration model comparison} traces the counterfactual from both models over a range of the parameter. At the point estimate, both models yield the same value for the counterfactual. Once we account for uncertainty in our estimates of the true parameter, deciding which policy to adopt becomes less straightforward: for higher values of $\theta$, the first policy is preferred, while the opposite is true for lower values.%---------------------------------------------------------------------------------------------------
%---------------------------------------------------------------------------------------------------
\subsection{Uncertainty sets}\label{Uncertainty set}
%---------------------------------------------------------------------------------------------------
%---------------------------------------------------------------------------------------------------
\citet{Manski.2021} suggests acknowledging parametric uncertainty by working with estimated confidence sets instead of point estimates. A confidence set $\bTheta(\alpha) \subset \bTheta$ covers the true parameters, from an ex-ante point of view, with a predetermined coverage probability of $(1 - \alpha)$. Proceeding with our analysis, we refine the status quo procedure, in which estimated parameter values serve as a stand-in for the model's true parametrization. Instead, we assume the estimated confidence set for the parameters  $\hat{\bTheta}(\alpha)$ and the counterfactual $\hat{\bTheta}_{y_g}(\alpha)$ are correct and analyze policy decisions accordingly. \\

\noindent Based on the estimated confidence sets, we construct so-called uncertainty sets for the parameters $\U(\alpha)$ and the prediction $\U_{y_g}(\alpha)$ by only considering parameterizations that we cannot reject based on a hypothesis test with confidence level $1 - \alpha$. This approach ensures the tractability of our decision-theoretic analysis, as the uncertainty set of the parameters is much smaller than the whole parameter space of the model. We adopt this procedure from the literature on data-driven robust optimization in operations research \citep{Ben-Tal.2013,Bertsimas.2018}.
%---------------------------------------------------------------------------------------------------
%---------------------------------------------------------------------------------------------------
\subsection{Statistical decision theory}\label{Statistical decision theory}
%---------------------------------------------------------------------------------------------------
%---------------------------------------------------------------------------------------------------
\noindent In our setting, a policy-maker relies on a structural model with an uncertain parametrization to map alternative policies to counterfactual predictions. In most cases, the preferred policy depends on the model's uncertain true parameters. We, therefore, draw on statistical decision theory to organize the decision-making process \citep{Gilboa.2009,Marinacci.2015}.\\

\noindent Returning to our example, we rank the two policies according to alternative statistical decision rules using an uncertainty set derived from a confidence set with a $90\%$ coverage probability. In what follows, we postulate a simple linear utility function $U(y_g)$ to describe the policy-maker's preferences.\footnote{We assume that the sampling distribution of the point estimate is normal with a mean of three and a standard deviation of three-fourths. We can derive the uncertainty sets directly and simply consider realizations of $\theta\in[1.76, 4.23]$.}\\

\begin{figure}[h!]\centering
\scalebox{0.35}{\includegraphics{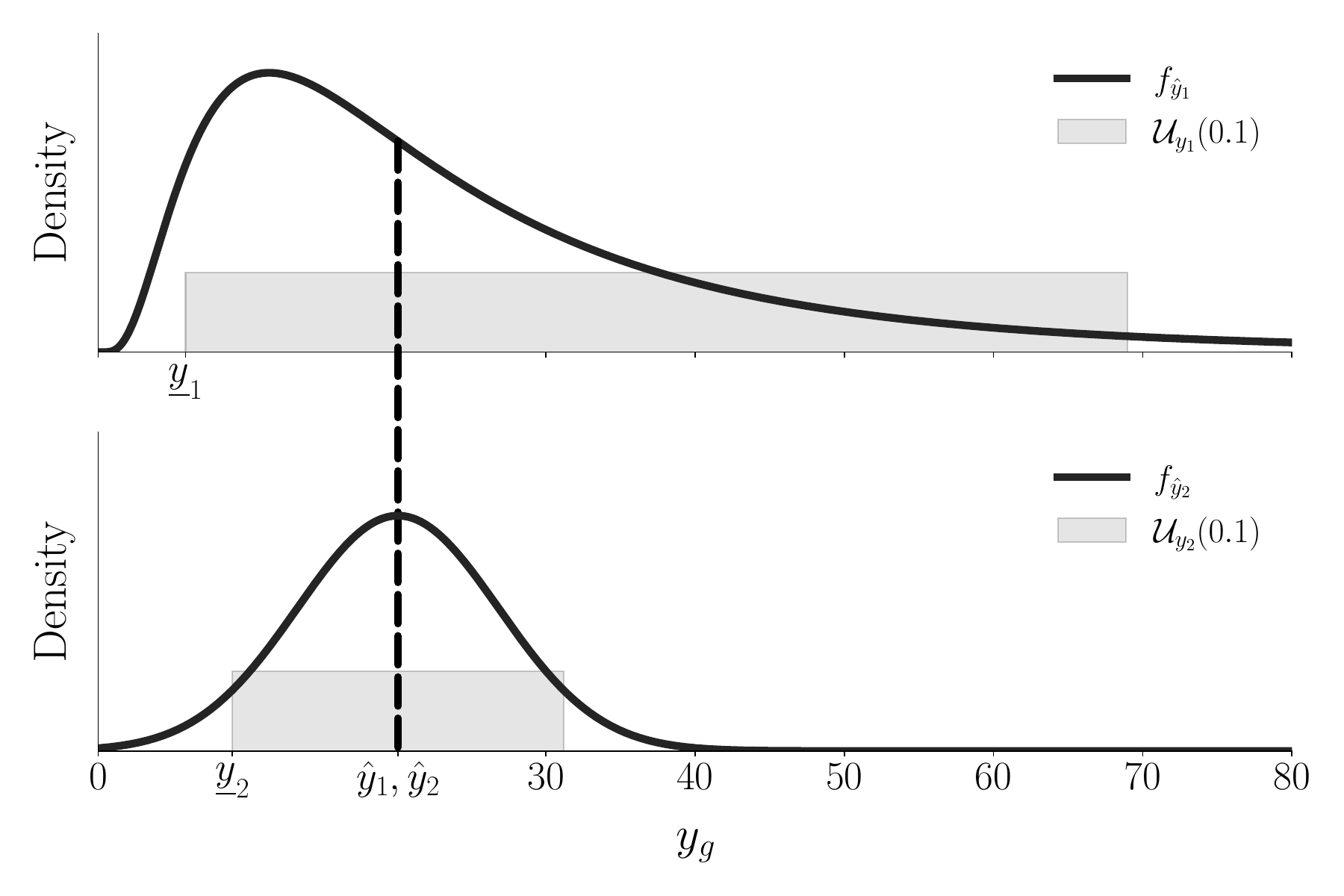}}
\caption{Comparing policy predictions}\label{Comparing policy predictions}
\end{figure}

\noindent Figure \ref{Comparing policy predictions} shows the implied sampling distribution of the predictions for the two alternative policies and the corresponding uncertainty sets $\U_{y_g}(0.1)$. The mapping $\M_1$ is highly nonlinear, while the mapping $\M_{2}$ is linear. When evaluated at the point estimate, the counterfactual is the same under both policies, so a policy-maker is indifferent. However, the spread of the uncertainty set differs considerably.\\

\FloatBarrier

\noindent Decision theory proposes a variety of different rules for reasonable decisions in this setting. We explore the following four: (1) as-if optimization, (2) maximin criterion, (3) minimax regret rule, and (4) subjective Bayes.\\

\noindent As-if optimization describes the predominant practice. The estimation of the model produces point estimates that serve as a plug-in for the true parameters. The decision maximizes the utility at the point estimate. More formally,
\begin{align*}
  g^* =\argmax_{g \in \G} U(M_g(\hat{\btheta})).
\end{align*}
Given our example, an as-if policy-maker is indifferent between the two policies, since both policies result in the same counterfactual at the point estimates as indicated by the dashed line in Figure \ref{Comparing policy predictions}.\\

\noindent The maximin criterion and minimax regret rule are two common alternatives that favor actions that work uniformly well over all possible parameters in the uncertainty set. This approach departs from as-if optimization, which only considers a policy's performance at a single point in the uncertainty set. The maximin decision \citep{Gilboa.1989, Wald.1950} is determined by computing the minimum utility for each policy within the uncertainty set and choosing the one with the highest worst-case outcome. Stated concisely,
\begin{align*}
 g^* =\argmax_{g \in \G}\min_{\btheta \in \U(\alpha)} U(M_g(\btheta)).
\end{align*}
\noindent Returning to Figure \ref{Comparing policy predictions}, a maximin policy-maker prefers $g_2$ as the worst-case outcome. Within the uncertainty set, $\underline{y}_2$ is better than under the alternative policy, $g_1$.\\

\begin{figure}[h!]\centering
\scalebox{0.35}{\includegraphics{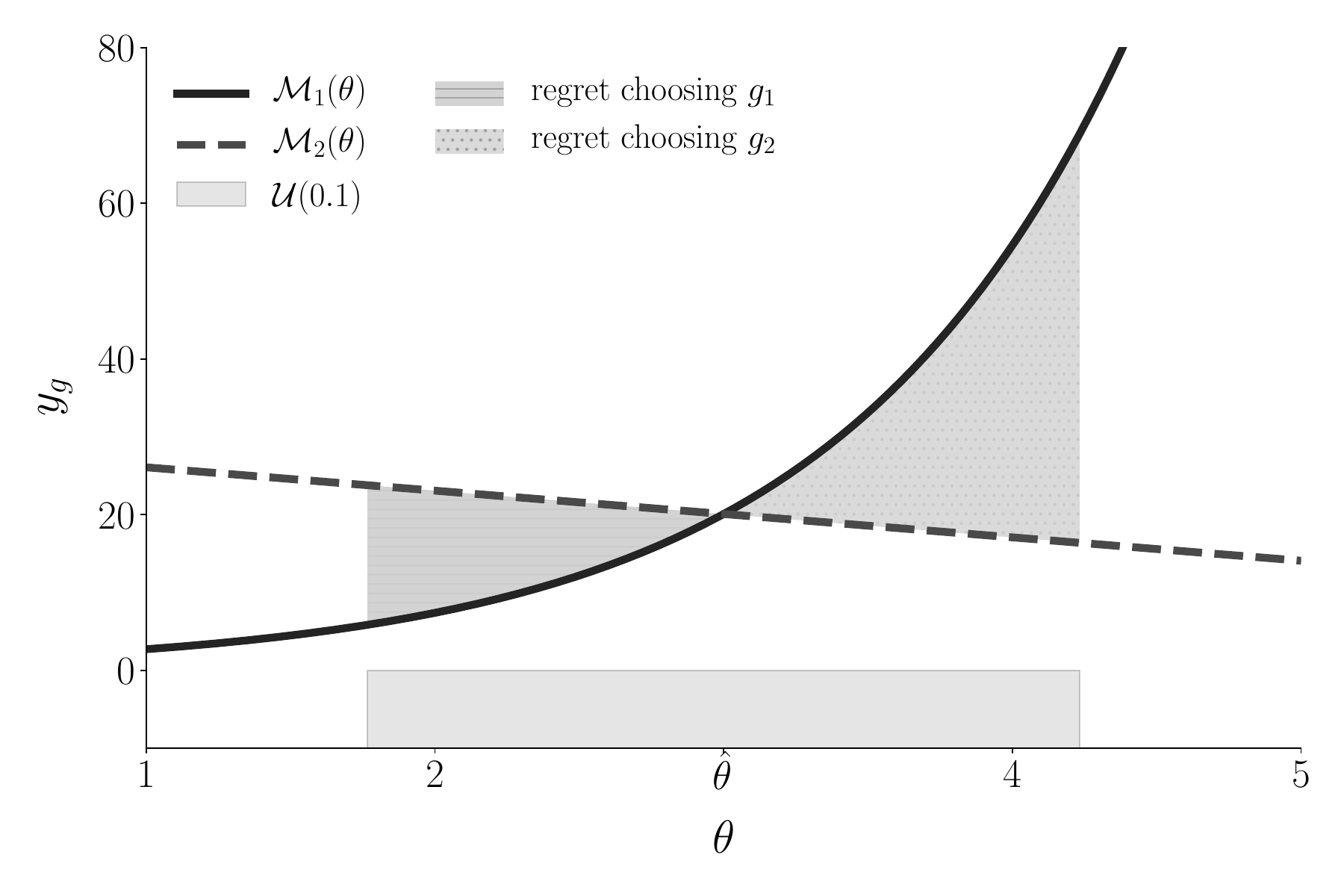}}
\caption{Comparing policy regret}\label{Comparing policy regret}\vspace{0.5cm}
\end{figure}

\noindent The minimax regret rule \citep{Manski.2004, Niehans.1948} computes the maximum regret for each policy over the whole uncertainty set and chooses the policy that minimizes the maximum regret. The regret of choosing a policy $g$ for a given parameterization of the model is the difference between the maximum possible utility achieved from adopting $\tilde{g} \in \G$ and the actual utility obtained. The decision maximizes:
\begin{align*}
  g^* =\argmin_{g \in \G} \max_{\btheta \in \U(\alpha)}  \underbrace{\left[\max_{\tilde{g} \in \G} U(M_{\tilde{g}}(\btheta)) - U(M_g(\btheta)) \right]}_{\text{regret}}.
\end{align*}
\noindent Figure \ref{Comparing policy regret} compares our two policy examples over the uncertainty sets. A policy-maker adopting policy $g_1$ regrets his choice for small values of the model parameter, while the opposite is true for larger values. The regret of each policy is maximized at the boundaries of the uncertainty set. Maximum regret is minimized when a policy-maker chooses $g_1$. It corresponds to the difference in the counterfactual at the lower boundary of the uncertainty set instead of the larger difference at its upper bound. This outcome contradicts the maximin decision in which policy $g_2$ is preferred.

\FloatBarrier

\noindent Each decision rule presented so far focuses on a single point in the uncertainty set as the policy's relevant performance measure. Bayesian approaches aggregate a policy's performance over the complete uncertainty set.\\

\noindent Maximization of the subjective expected utility \citep{Savage.1954} requires the policy-maker to place a subjective probability distribution $f_{\btheta}$ over the parameters in the uncertainty set.  A policy-maker then selects the alternative with the highest expected subjective utility. Formally,
\begin{align*}
  g^* =\argmax_{g \in \G} \int_{\U(\alpha)} U(M_g({\btheta})) \diff f_{\btheta}.
\end{align*}
\noindent Applying a uniform distribution to our example, a policy-maker chooses $g_1$, which performs well for high values of $\btheta$ and still reasonably well for low values.

%!TEX root = ../main.tex
%---------------------------------------------------------------------------------------------------
\FloatBarrier\section{Eckstein-Keane-Wolpin models}\label{Setting}
%---------------------------------------------------------------------------------------------------
We now present the general structure of Eckstein-Keane-Wolpin (EKW) models \citep{Aguirregabiria.2010}  and their solution approach. We then turn to the customized version used by \citet{Keane.1997} to study the career decisions of young men and investigate the consequences of parametric uncertainty in this empirically-grounded and computationally demanding setting. We outline their model's basic setup, provide some descriptive statistics of the empirical data used in our estimation, and then discuss the core findings.
%---------------------------------------------------------------------------------------------------
%---------------------------------------------------------------------------------------------------
\subsection{General structure}
%---------------------------------------------------------------------------------------------------
%---------------------------------------------------------------------------------------------------
EKW models describe sequential decision-making under uncertainty \citep{Gilboa.2009, Machina.2014}. At time $t = 1, \hdots, T$ each individual observes the state of their choice environment $s_t\in S$ and chooses an action $a_t$ from the set of admissible actions $\mathcal{A}$. The decision has two consequences: an individual receives an immediate utility $u_t(s_t, a_t)$ and their environment evolves to a new state $s_{t + 1}$. The transition from $s_t$ to $s_{t + 1}$ is affected by the action but remains uncertain. Since individuals are forward-looking, they do not simply choose the alternative with the highest immediate utility. Instead, they take the future consequences of their actions into account.\\

\noindent A policy $\pi =(d^\pi_1, \hdots, d^\pi_T)$ provides the individual with instructions for choosing an action in any possible future state. It is a sequence of decision rules $d^\pi_t$ that specify the action $d^\pi_t(s_t) \in \mathcal{A}$   at a particular time $t$ for any possible state $s_t$ under $\pi$. The implementation of a policy generates a sequence of utilities that depends on the objective transition probability distribution $p_t(s_t, a_t)$ for the evolution from state $s_t$ to $s_{t + 1}$ induced by the model.\\

\begin{figure}\centering
	%!TEX root = ../main.tex

\definecolor{light-gray}{gray}{0.85}

\begin{tikzpicture}[node distance=2cm]
%define styles
\tikzstyle{startstop} = [circle, rounded corners, minimum width=0.6cm, minimum height=0.3cm,text centered, draw=black]
[
->,
>=stealth',
auto,node distance=3cm,
thick,
main node/.style={circle, draw, font=\sffamily\Large\bfseries}
]
\tikzstyle{arrow} = [thick,->,>=stealth]]
\tikzstyle{darrow} = [dotted,->,>=stealth]]
%first and second column

\node (r0) [startstop, xshift = -3cm, draw = none] {};
\node (r999) [startstop, xshift = 13cm, draw = none] {};

\node (r1) [startstop, xshift = -1cm] {\footnotesize $~\,s_t\,~$};
\node (r2) [startstop, xshift = 5cm] {\footnotesize $s_{t+1}$};  %previously 4
\node (r3) [startstop, xshift = 11cm] {\footnotesize $s_{t+2}$}; %previously 8

\draw [arrow, dashed] (r0) -- node[anchor=south] {} (r1) ;
\draw [arrow] (r1) -- node[anchor=south] {} (r2) ;
\draw [arrow] (r2) -- node[anchor=south] {} (r3) ;
\draw [arrow, dashed] (r3) -- node[anchor=south] {} (r999) ;

\node (r4) [startstop, xshift = 0 cm, yshift = -2.5cm, inner sep = 0.08cm] {\footnotesize $~\,a_t\,~$ };
\node (r5) [startstop, xshift = 3 cm, yshift = -2.5cm, inner sep = 0.08cm] {\footnotesize $~\,u_t\,~$ };
\node (r6) [startstop, xshift = 6 cm, yshift = -2.5cm, inner sep = 0.08cm] {\footnotesize $a_{t+1}$ };
\node (r7) [startstop, xshift = 9 cm, yshift = -2.5cm, inner sep = 0.08cm] {\footnotesize $u_{t+1}$ };

\draw [arrow, color = gray] (r1) -- node[anchor=south] {} (r4) ;
\draw [arrow, color = gray] (r2) -- node[anchor=south] {} (r6) ;

\draw[ thick](0,-2.05).. controls (0.75, -0.2) and (0.8,0)..(1.5, 0);
\draw[->, thick](1.5, 0).. controls (2.15, 0) and (2.25, -0.2)..(3, -2.05);
\draw[ thick](6,-2.05).. controls (6.75, -0.2) and (6.85,0)..(7.5, 0);
\draw[->, thick](7.5, 0).. controls (8.15, 0) and (8.25,-0.2)..(9, -2.05);

\node(r8)[startstop, xshift = - 1.75cm, yshift = -1.3cm, draw =none, color = gray, align=center] {\footnotesize decide \\ \footnotesize $d_t^{\pi}(s_t)$ };
\node(r9)[startstop, xshift= 4.2cm, yshift = -1.3cm, draw =none, color =gray, align = center] {\footnotesize decide \\ \footnotesize $d_{t+1}^{\pi}(s_{t+1})$ };
\node(r10)[startstop, yshift = 1cm, xshift = 2cm, draw =none, align=center ] {\footnotesize transition \\ \footnotesize $p_t(s_t, a_t)$};
\node(r11)[startstop, yshift = 1cm, xshift = 8cm, draw =none, align=center ] { \footnotesize transition \\ \footnotesize $p_{t+1}(s_{t+1}, a_{t+1})$};
\node(r12)[startstop, yshift = -1.3cm, xshift = 1.5cm, draw =none, align=center ] {\footnotesize receive\\ \footnotesize  $u_t(s_t, a_t)$ };
\node(r13)[startstop, yshift = -1.3cm, xshift = 7.5cm, draw =none, align=center ] {\footnotesize receive\\ \footnotesize  $u_{t+1}(s_{t+1}, a_{t+1})$ };

\end{tikzpicture}
\caption{Timing of events}\label{Timing}
\end{figure}
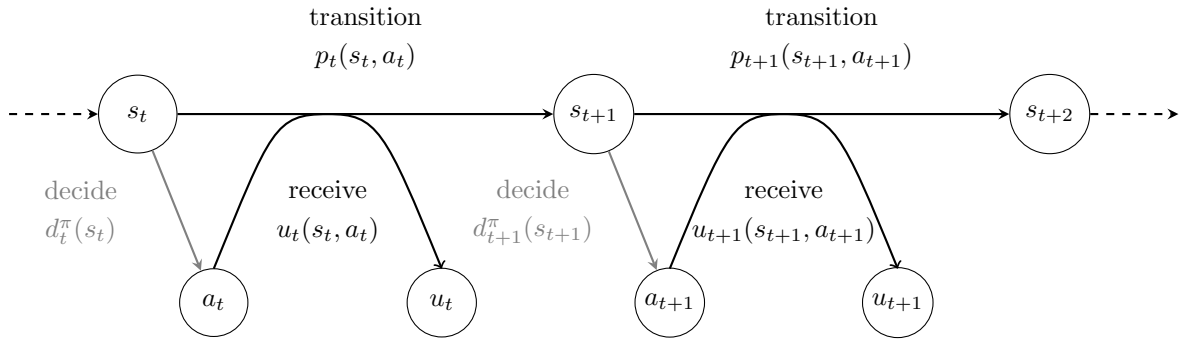

\noindent Figure \ref{Timing} depicts the timing of events for two generic periods.  At the beginning of period $t$, an individual fully learns about each action's immediate utility, selects one of the alternatives, and receives its immediate utility. Then, the state evolves from $s_t$ to $s_{t + 1}$, and the process repeats itself in $t + 1$.\\

\noindent Individuals make their decisions facing uncertainty about the future and seek to maximize their expected total discounted utilities over all decision periods given all available information. They have rational expectations \citep{Muth.1961}, so their subjective beliefs about the future agree with the objective probabilities for all possible future events provided by the model. Immediate utilities are separable between periods \citep{Kahneman.1997}, and a discount factor $\delta$ parameterizes a preference for immediate over future utilities \citep{Samuelson.1937}.\\

\noindent Equation (\ref{Objective Risk}) formally describes the individual's objective. Given an initial state $s_1$, they implement a policy $\pi$ that maximizes the expected total discounted utilities over all decision periods given the information available at the time.
\begin{align}\label{Objective Risk}
\max_{\pi \in \Pi} \E_{s_1}^\pi\left[\,\sum^{T}_{t = 1}  \delta^{t - 1} u_t(s_t, d^\pi_t(s_t))\,\right]
\end{align}

\noindent EKW models are set up as a standard Markov decision process (MDP) \citep{Puterman.1994,Rust.1994,White.1993} that can be solved by a simple backward induction procedure. In the final period $T$, there is no future to consider, and the optimal action is choosing the alternative with the highest immediate utility in each state. With the decision rule for the final period, we can determine all other optimal decisions recursively. We use our group's open-source research code \verb+respy+ \citep{Gabler.2020b}, which allows for the flexible specification, simulation, and estimation of EKW models. Detailed documentation of the software and its numerical components is available at \url{http://respy.readthedocs.io}.
%---------------------------------------------------------------------------------------------------
%---------------------------------------------------------------------------------------------------
\subsection{The career decisions of young men}
%---------------------------------------------------------------------------------------------------
%---------------------------------------------------------------------------------------------------
\citet{Keane.1997} specialize the model above to explore the career decisions of young men regarding their schooling, work, and occupational choices using the National Longitudinal Survey of Youth 1979 (NLSY79) \citep{NLSY.2019} for the estimation of the model. We restrict ourselves to a basic summary of their setup. Further documentation of the model specification and the observed dataset is available in the Appendix.\\

\noindent \citet{Keane.1997} follows individuals over their working life from young adulthood at age 16 to retirement at age 65. Each decision period $t = 16, \dots, 65$ represents a school year. Figure \ref{Decision tree} illustrates the initial decision problem as individuals select one of five alternatives from the set of admissible actions $a\in\mathcal{A}$. They can  decide to either work in a blue-collar or a white-collar occupation ($a = 1, 2$), serve in the military $(a = 3)$, attend school $(a = 4)$, or stay at home $(a = 5)$.\\

\begin{figure}[t!]\centering
	\scalebox{0.75}{% Decision tree black white

\tikzset{
	treenode/.style = {shape=rectangle, rounded corners, draw, align=center, bottom color=blue!20},
	root/.style     = {treenode, font=\small, draw=none},
	env/.style      = {treenode, font=\small, draw=none},
	dummy/.style    = {circle,draw}
}

\begin{tikzpicture}
[
	x=30pt,
	y=26pt,
	yscale=-1,
	xscale=1,
	baseline=-120pt,
	grow                    = right,
	edge from parent/.style = {draw, -latex},
	every node/.style       = {font=\footnotesize, minimum width={width("Magnetometer")+2pt}},
	sloped
]

% Zero  level: START
\node [root, top color = white, bottom color=white, draw] (0) at (-10,0) {\textbf{Start}};

% First level: BLUE
\node [env, top color = bwtabblue, bottom color = bwtabblue, color = bwtaborange, scale = 0.8] (1) at (-5,-4.2) {\textbf{Blue}};
\draw[->, thick] (0) edge node[left of = 0, rotate = 72.75, node distance = 0.3cm]{} (1);
% Second level: Blue, White, Military, School, Home
\node [env, top color = bwtabblue, bottom color = bwtabblue, color = bwtaborange, scale = 0.55] (11) at (0,-5) {Blue};
\draw[->] (1) edge (11);
\node [env, top color = bwtabred, bottom color = bwtabred, color = bwtaborange, scale = 0.55] (12) at (0,-4.6) {White};
\draw[->] (1) edge (12);
\node [env, top color = bwtabpurple, bottom color = bwtabpurple, scale = 0.55] (13) at (0,-4.2) {Military};
\draw[->] (1) edge (13);
\node [env, top color = bwtaborange, bottom color = bwtaborange, scale = 0.55] (14)  at (0,-3.8) {School};
\draw[->] (1) edge (14);
\node [env, top color = bwtabgreen, bottom color = bwtabgreen, scale = 0.55] (15)  at (0,-3.4) {Home};
\draw[->] (1) edge (15);
% Third level coordinates
\coordinate (a1) at (3,-5);
\draw [->, dashed, color=tabgrey] (11) to[right] node[auto] {} (a1);
\coordinate (a2) at (3,-4.6);
\draw [->, dashed, color=tabgrey] (12) to[right] node[auto] {} (a2);
\coordinate (a3) at (3,-4.2);
\draw [->, dashed, color=tabgrey] (13) to[right] node[auto] {} (a3);
\coordinate (a4) at (3,-3.8);
\draw [->, dashed, color=tabgrey] (14) to[right] node[auto] {} (a4);
\coordinate (a5) at (3,-3.4);
\draw [->, dashed, color=tabgrey] (15) to[right] node[auto] {} (a5);

%First level: WHITE
\node [env, top color = bwtabred, bottom color = bwtabred, color = bwtaborange, scale=0.8] (2) at (-5,-2.1) {\textbf{White}};
\draw[->, thick] (0) edge node[right of = 0, yshift=0.35cm,  rotate = 41.0, node distance = 0cm]{} (2);
% Second level: Blue, White, Military, School, Home
\node [env, top color = bwtabblue, bottom color = bwtabblue, color = bwtaborange, scale = 0.55] (21) at (0,-2.9) {Blue};
\draw[->] (2) edge (21);
\node [env, top color = bwtabred, bottom color = bwtabred, color = bwtaborange, scale = 0.55] (22) at (0,-2.5) {White};
\draw[->] (2) edge (22);
\node [env, top color = bwtabpurple, bottom color = bwtabpurple, scale = 0.55] (23) at (0,-2.1) {Military};
\draw[->] (2) edge (23);
\node [env, top color = bwtaborange, bottom color = bwtaborange, scale = 0.55] (24)  at (0,-1.7) {School};
\draw[->] (2) edge (24);
\node [env, top color = bwtabgreen, bottom color = bwtabgreen, scale = 0.55] (25)  at (0,-1.3) {Home};
\draw[->] (2) edge (25);
% Second level --> third level
\coordinate (e1) at (3,-2.9);
\draw [->, dashed, color = tabgrey] (21) to[right] node[auto] {} (e1);
\coordinate (e2) at (3,-2.5);
\draw [->, dashed, color = tabgrey] (22) to[right] node[auto] {} (e2);
\coordinate (e3) at (3,-2.1);
\draw [->, dashed, color = tabgrey] (23) to[right] node[auto] {} (e3);
\coordinate (e4) at (3,-1.7);
\draw [->, dashed, color = tabgrey] (24) to[right] node[auto] {} (e4);
\coordinate (e5) at (3,-1.3);
\draw [->, dashed, color = tabgrey] (25) to[right] node[auto] {} (e5);

% First Level: MILITARY
\node [env, top color = bwtabpurple, bottom color = bwtabpurple, scale = 0.8] (3) at (-5,0) {\textbf{Military}};
\draw[->, thick] (0) edge node[right of = 0, yshift = 0.25cm, node distance = 0.2cm]{} (3);
% Second level: Blue, White, Military, School, Home
\node [env, top color = bwtabblue, bottom color = bwtabblue, color = bwtaborange, scale = 0.55] (31) at (0,-0.8) {Blue};
\draw[->] (3) edge (31);
\node [env, top color = bwtabred, bottom color = bwtabred, color = bwtaborange, scale = 0.55] (32) at (0,-0.4) {White};
\draw[->] (3) edge (32);
\node [env, top color = bwtabpurple, bottom color = bwtabpurple, scale = 0.55] (33) at (0,0) {Military};
\draw[->] (3) edge (33);
\node [env, top color = bwtaborange, bottom color = bwtaborange, scale = 0.55] (34)  at (0,0.4) {School};
\draw[->] (3) edge (34);
\node [env, top color = bwtabgreen, bottom color = bwtabgreen, scale = 0.55] (35)  at (0,0.8) {Home};
\draw[->] (3) edge (35);
% Second level --> third level
\coordinate (c1) at (3,-0.8);
\draw [->, dashed, color = tabgrey] (31) to[right] node[auto] {} (c1);
\coordinate (c2) at (3,-0.4);
\draw [->, dashed, color = tabgrey] (32) to[right] node[auto] {} (c2);
\coordinate (c3) at (3,0);
\draw [->, dashed, color = tabgrey] (33) to[right] node[auto] {} (c3);
\coordinate (c4) at (3,0.4);
\draw [->, dashed, color = tabgrey] (34) to[right] node[auto] {} (c4);
\coordinate (c5) at (3,0.8);
\draw [->, dashed, color = tabgrey] (35) to[right] node[auto] {} (c5);

% First Level: SCHOOL
\node [env, top color = bwtaborange, bottom color = bwtaborange, scale = 0.8] (4)  at (-5,2.1) {\textbf{School}};
\draw[->, thick] (0) edge node[right of = 0, yshift=0.1cm, rotate = -38, node distance = 0.25cm]{} (4);
% Second level: Blue, White, Military, School, Home
\node [env, top color = bwtabblue, bottom color=bwtabblue, color = bwtaborange, scale = 0.55] (41) at (0,1.3) {Blue};
\draw[->] (4) edge (41);
\node [env, top color = bwtabred, bottom color=bwtabred, color = bwtaborange, scale = 0.55] (42) at (0,1.7) {White};
\draw[->] (4) edge (42);
\node [env, top color = bwtabpurple, bottom color=bwtabpurple, scale = 0.55] (43) at (0,2.1) {Military};
\draw[->] (4) edge (43);
\node [env, top color = bwtaborange, bottom color=bwtaborange, scale = 0.55] (44)  at (0,2.5) {School};
\draw[->] (4) edge (44);
\node [env, top color = bwtabgreen, bottom color=bwtabgreen, scale = 0.55] (45)  at (0,2.9) {Home};
\draw[->] (4) edge (45);
% Second level --> third level
\coordinate (b1) at (3,1.3);
\draw [->, dashed, color = tabgrey] (41) to[right] node[auto] {} (b1);
\coordinate (b2) at (3,1.7);
\draw [->, dashed, color = tabgrey] (42) to[right] node[auto] {} (b2);
\coordinate (b3) at (3,2.1);
\draw [->, dashed, color = tabgrey] (43) to[right] node[auto] {} (b3);
\coordinate (b4) at (3,2.5);
\draw [->, dashed, color = tabgrey] (44) to[right] node[auto] {} (b4);
\coordinate (b5) at (3,2.9);
\draw [->, dashed, color = tabgrey] (45) to[right] node[auto] {} (b5);

% First Level: HOME
\node [env, top color = bwtabgreen, bottom color = bwtabgreen, scale=0.8] (5)  at (-5,4.2) {\textbf{Home}};
\draw[->, thick] (0) edge node[right of = 0, yshift=0.25cm, rotate = -72.75, node distance = 0.15cm]{} (5);
%\draw[->, thick] (0) edge (5);
% Second level: Blue, White, Military, School, Home
\node [env, top color = bwtabblue, bottom color = bwtabblue, color = bwtaborange, scale = 0.55] (51)  at (0,3.4) {Blue};
\draw[->] (5) edge (51);
\node [env, top color = bwtabred, bottom color = bwtabred, color = bwtaborange, scale = 0.55] (52)  at (0,3.8) {White};
\draw[->] (5) edge (52);
\node [env, top color = bwtabpurple, bottom color = bwtabpurple, scale = 0.55] (53) at (0,4.2) {Military};
\draw[->] (5) edge (53);
\node [env, top color = bwtaborange, bottom color = bwtaborange, scale = 0.55] (54) at (0,4.6) {School};
\draw[->] (5) edge (54);
\node [env, top color = bwtabgreen, bottom color = bwtabgreen, scale=0.55] (55) at (0,5.0) {Home};
\draw[->] (5) edge (55);
% Second level --> third level
\coordinate (d1) at (3,3.4);
\draw [->, dashed, color = tabgrey] (51) to[right] node[auto] {} (d1);
\coordinate (d2) at (3,3.8);
\draw [->, dashed, color = tabgrey] (52) to[right] node[auto] {} (d2);
\coordinate (d3) at (3,4.2);
\draw [->, dashed, color = tabgrey] (53) to[right] node[auto] {} (d3);
\coordinate (d4) at (3,4.6);
\draw [->, dashed, color = tabgrey] (54) to[right] node[auto] {} (d4);
\coordinate (d5) at (3,5.0);
\draw [->, dashed, color = tabgrey] (55) to[right] node[auto] {} (d5);

\end{tikzpicture}

\vspace{1cm}}
	\caption{Decision tree}\label{Decision tree}
\end{figure}
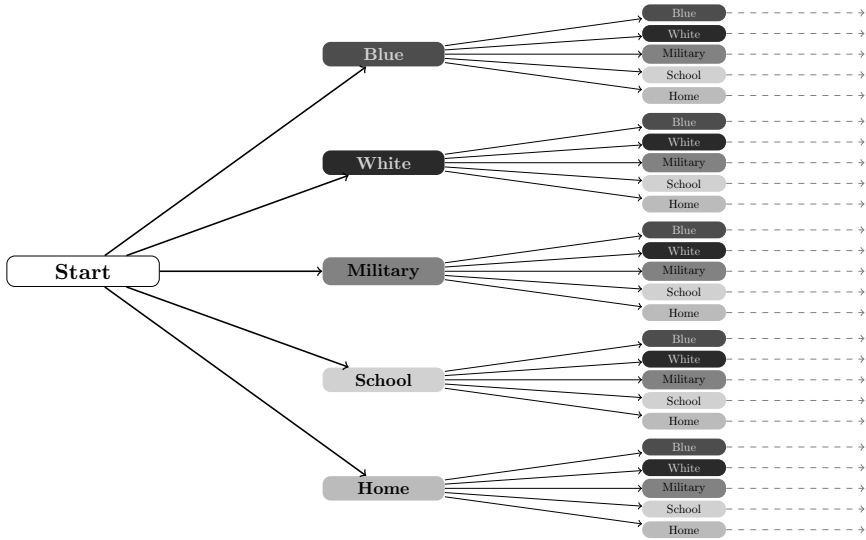\FloatBarrier

\noindent Individuals are already heterogeneous when entering the model. They differ with respect to their level of initial schooling $h_{16}$, and have one of four different $\mathcal{J} = \{1, \hdots, 4\}$ alternative-specific skill endowment types $\bm{e} = \left(e_{j,a}\right)_{\mathcal{J} \times \mathcal{A}}$.\\

\noindent The immediate utility $u_a(\cdot)$ of each alternative consists of a non-pecuniary utility $\zeta_a(\cdot)$ and, at least for the working alternatives, an additional wage component $w_a(\cdot)$. Both depend on the level of human capital as measured by their alternative-specific skill endowment $\bm{e}$, their years of completed schooling $h_t$, and their occupation-specific work experience $\bm{k_t} = \left(k_{a,t}\right)_{a\in\{1, 2, 3\}}$. The immediate utilities are influenced by last-period choices $a_{t -1}$ and alternative-specific productivity shocks $\bm{\epsilon_t} = \left(\epsilon_{a,t}\right)_{a\in\mathcal{A}}$ as well. Their general form is given by:
\begin{align*}
u_a(\cdot) =
\begin{cases}
    \zeta_a(\bm{k_t}, h_t, t, a_{t -1})  + w_a(\bm{k_t}, h_t, t, a_{t -1}, e_{j, a}, \epsilon_{a,t})                & \text{if}\, a \in \{1, 2, 3\}  \\
    \zeta_a(\bm{k_t}, h_t, t, a_{t-1}, e_{j,a}, \epsilon_{a,t})                                                  &  \text{if}\, a \in \{4, 5\}.
\end{cases}
\end{align*}
\noindent Work experience $\bm{k_t}$  and years of completed schooling $h_t$ evolve deterministically. There is no uncertainty about grade completion \citep{Altonji.1993} and no part-time enrollment. Schooling is defined by time spent in school, not by formal credentials acquired. Once individuals reach a certain amount of schooling, they acquire a degree.
\begin{align*}
k_{a,t+1} = k_{a,t} + \ind[a_t = a]  &\qquad \text{if}\, a \in \{1, 2, 3\} \\
h_{t + 1\phantom{,a}} = h_{t\phantom{,a}} +   \ind[a_t = 4]  &\qquad
\end{align*}
\noindent The productivity shocks $\bm{\epsilon_t}$ are uncorrelated across time and follow a multivariate normal distribution with mean $\bm{0}$ and covariance matrix $\bm{\Sigma}$. Given the structure of the utility functions and the distribution of the shocks, the state at time $t$ is $s_t = \{\bm{k_t}, h_t, t, a_{t -1}, \bm{e},\bm{\epsilon_t}\}$.\\

\noindent Skill endowments $\bm{e}$ and initial schooling $h_{16}$ are the only sources of persistent heterogeneity in the model. All remaining differences in life-cycle decisions result from different transitory shocks $\bm{\epsilon_t}$ that occur over time.\\

\noindent Theoretical and empirical research from specialized disciplines within economics informs the specification of each $u_a(\cdot)$. As an example, we provide the exact functional form of the non-pecuniary utility from schooling in Equation (\ref{eq:UtilitySchoolingMain}). Further details on the specification of the utility functions are available in the Appendix.
\begin{align}\label{eq:UtilitySchoolingMain}
		\zeta_4(s_t)   = & \underbrace{e_{j,4}}_{\text{type}} + \underbrace{\beta_{tc_1} \cdot \ind[h_t \geq 12] + \beta_{tc_2} \cdot \ind[h_t \geq 16]}_{\text{tuition costs}}  + \underbrace{\gamma_{4,4} \cdot t + \gamma_{4,5} \cdot \ind[t < 18]}_{\text{time trend}}  \\[15pt]
	    							  & + \underbrace{\beta_{rc_1} \cdot \ind[a_{t-1} \neq 4, h_t < 12]   + \beta_{rc_2} \cdot \ind[a_{t-1} \neq 4, h_t \geq 12]}_{\text{re-enrollment cost}}  + \hdots + \epsilon_{4,t}\nonumber
\end{align}
There is a direct cost in the form of tuition for continuing education after high school $\beta_{tc_1}$ and college $\beta_{tc_2}$. The decision to leave school is reversible, but entails re-enrollment costs that differ by schooling category ($\beta_{rc_1}, \beta_{rc_2}$).\\

\noindent We analyze the original dataset used by \citet{Keane.1997}. We only provide a brief description and relegate further details to the Appendix. The authors construct their sample based on the NLSY79, a nationally representative sample of young men and women living in the United States in 1979 and born between 1957 and 1964. Individuals were followed from 1979 onwards and repeatedly interviewed about their schooling decisions and labor market experiences. Based on this information, individuals are assigned to either working in one of the three occupations, attending school, or simply staying at home.\\

\noindent \citet{Keane.1997} restrict attention to white men, who turned 16 between 1977 and 1981, and exploit information collected between 1979 and 1987. Thus, individuals in the sample range in age between 16 and 26 years old. While the sample initially consists of 1,373 individuals at age 16, this number drops to 256 at the age of 26 due to sample attrition and missing data. Overall, the final sample consists of 12,359 person-period observations.\\

\begin{figure}[t!]\centering
\subfloat[Choices]{\scalebox{0.25}{\includegraphics{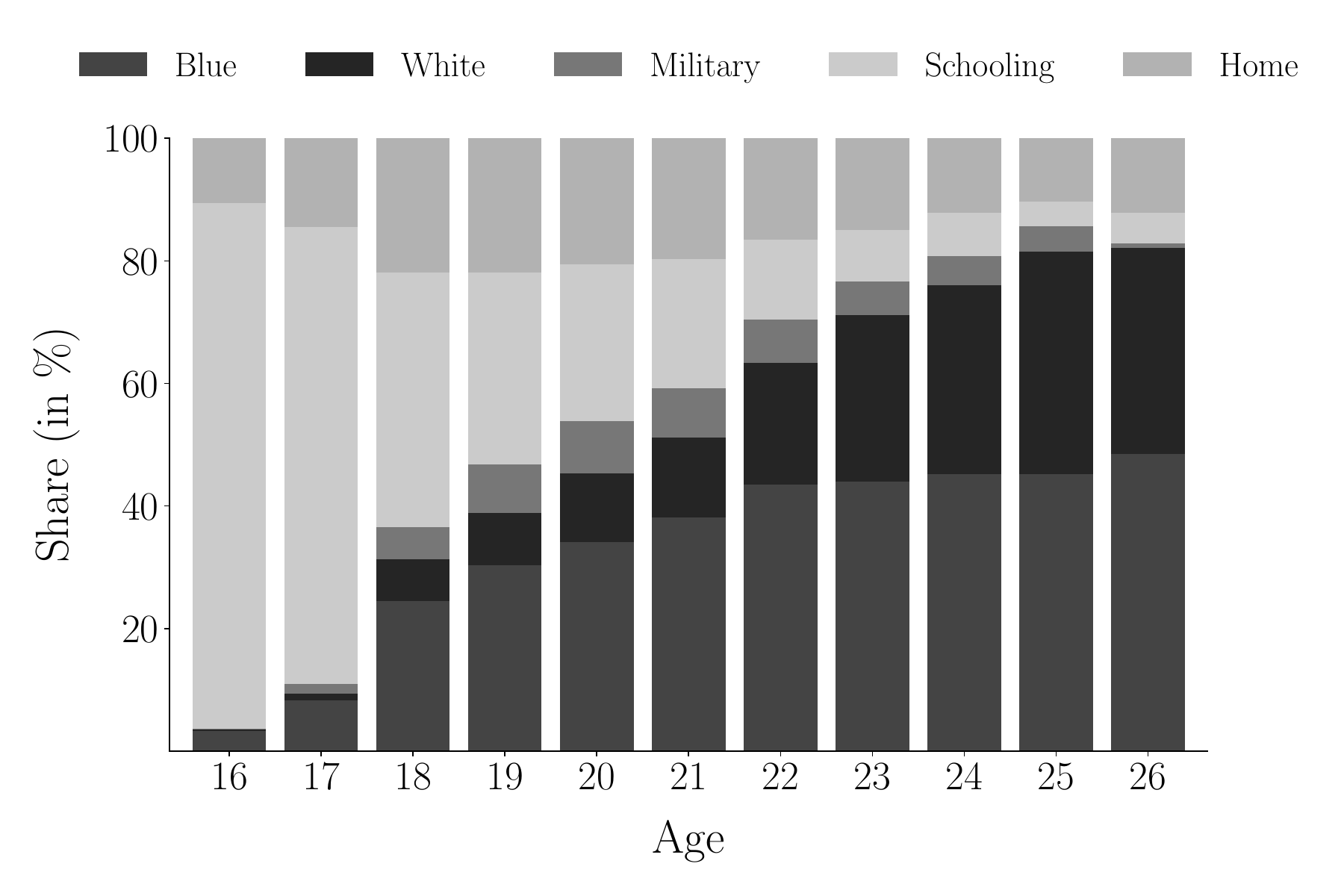}}}\hspace{0.3cm}
\subfloat[Average wage]{\scalebox{0.25}{\includegraphics{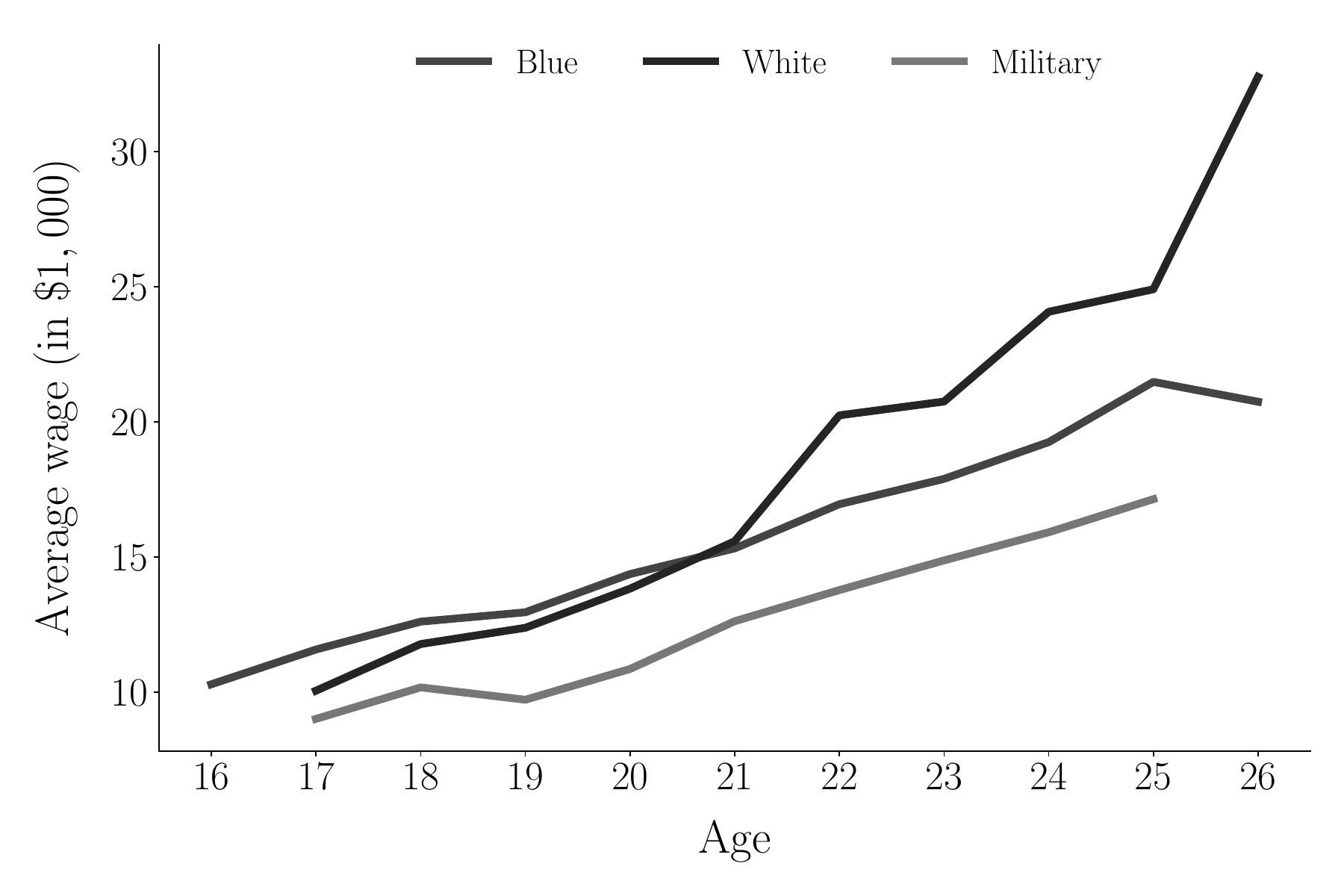}}}
\begin{center}
\begin{minipage}[t]{0.8\columnwidth}
\item \scriptsize{\textbf{Notes:} The wage is a full-time equivalent deflated by the gross national product deflator, with 1987 as the base year. We do not report the wage if less than ten observations are available.}
\end{minipage}
\end{center}
\caption{Data overview}\label{Overview}
\end{figure}

\noindent Figure \ref{Overview} summarizes the evolution of choices and wages over the sample period. Roughly 86\% of individuals initially enroll in school, but this share steadily declines with age. Nevertheless, about 39\% pursue some form of higher education and obtain more than a high school degree. As individuals leave school, most of them initially pursue a blue-collar occupation. However, the relative share of white-collar workers increases as individuals entering the labor market later gain access to higher levels of schooling. At age 26, about 48\% work in a blue-collar occupation and 34\% in a white-collar occupation. The share of individuals in the military peaks around age 20 at 8\%. At its maximum around age 18, approximately 20\% of individuals stay at home.\\

\FloatBarrier

\noindent For an individual, the average wage starts at about \$10,000 at age 16 and increases considerably up to about \$25,000 by the age of 26. While starting wages for blue-collar workers are about \$10,286, wages in white-collar occupations and the military start around \$9,000. However, wages for white-collar occupations increase sharply over time, overtaking blue-collar wages around age 21. By the end of the observation period, wages for white-collar occupations are about 50\% higher than blue-collar wages at \$32,756 compared to only \$20,739. Military wages remain lowest throughout.\\

\noindent We consider observations for $i = 1, \hdots, N$ individuals in each time period $t = 1, \dots, T_i$. For every observation $(i, t)$ in the data, we observe the action $a_{it}$, some components $\bar{u}_{it}$ of the utility, and a subset $\bar{s}_{it}$ of the state $s_{it}$. Therefore, from an economist's point of view, we must distinguish between two types of state variables $s_{it} = \{\bar{s}_{it}, \bm{e},\bm{\epsilon_t}\}$. At time $t$, the economist and individual both observe $\bar{s}_{it}$, while $\{ \bm{e},\bm{\epsilon_t}\}$ is only observed by the individual.\\

\noindent We use  simulated maximum likelihood \citep{Fisher.1922,Manski.1977} estimation and determine the $88$ model parameters $\hat{\btheta}$ that maximize the likelihood function $\mathcal{L}(\btheta\mid\mathcal{D})$. As we only observe a subset $\bar{s}_t = \{\bm{k_t}, h_t, t, a_{t -1}\}$ of the state, we can determine the probability $p_{it}(a_{it}, \bar{u}_{it} \mid \bar{s}_{it}, \btheta)$ of individual $i$ at time $t$ in $\bar{s}_{it}$ choosing $a_{it}$ and receiving $\bar{u}_{it}$ given parametric assumptions about the distribution of $\bm{\epsilon_t}$. The objective function takes the following form:
\begin{align*}
  \hat{\btheta} \equiv \argmax_{\btheta \in \bTheta}  \underbrace{\prod^N_{i= 1} \prod^{T_i}_{t= 1}\, p_{it}(a_{it}, \bar{u}_{it} \mid \bar{s}_{it}, \btheta)}_{\mathcal{L}(\btheta\mid\mathcal{D})}.
\end{align*}
\noindent Overall, our parameter estimates are in broad agreement with the results reported in the original paper and the related literature. For example, individuals discount future utilities by $6\%$ per year. The returns to schooling vary according to occupation. While wages for white-collar occupations increase by about $6\%$ with each additional year of schooling, they only increase by $2\%$ for those working blue collar jobs. Skills are transferable across occupations as work experience increases wages in both blue and white-collar occupations.\\

\begin{figure}[h!]\centering
\subfloat[Average wage]{\scalebox{0.25}{\includegraphics{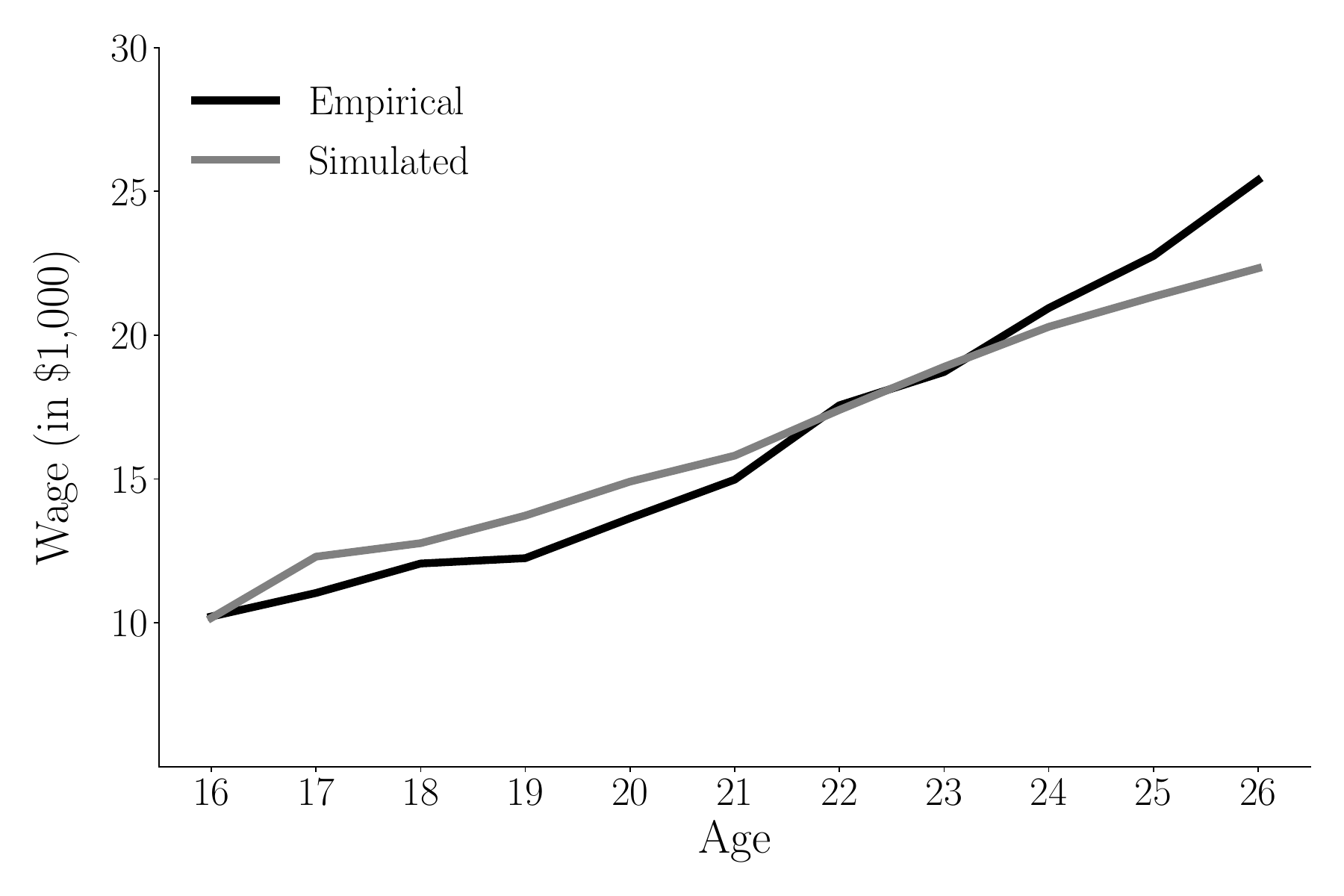}}}
\subfloat[Blue-collar]{\scalebox{0.25}{\includegraphics{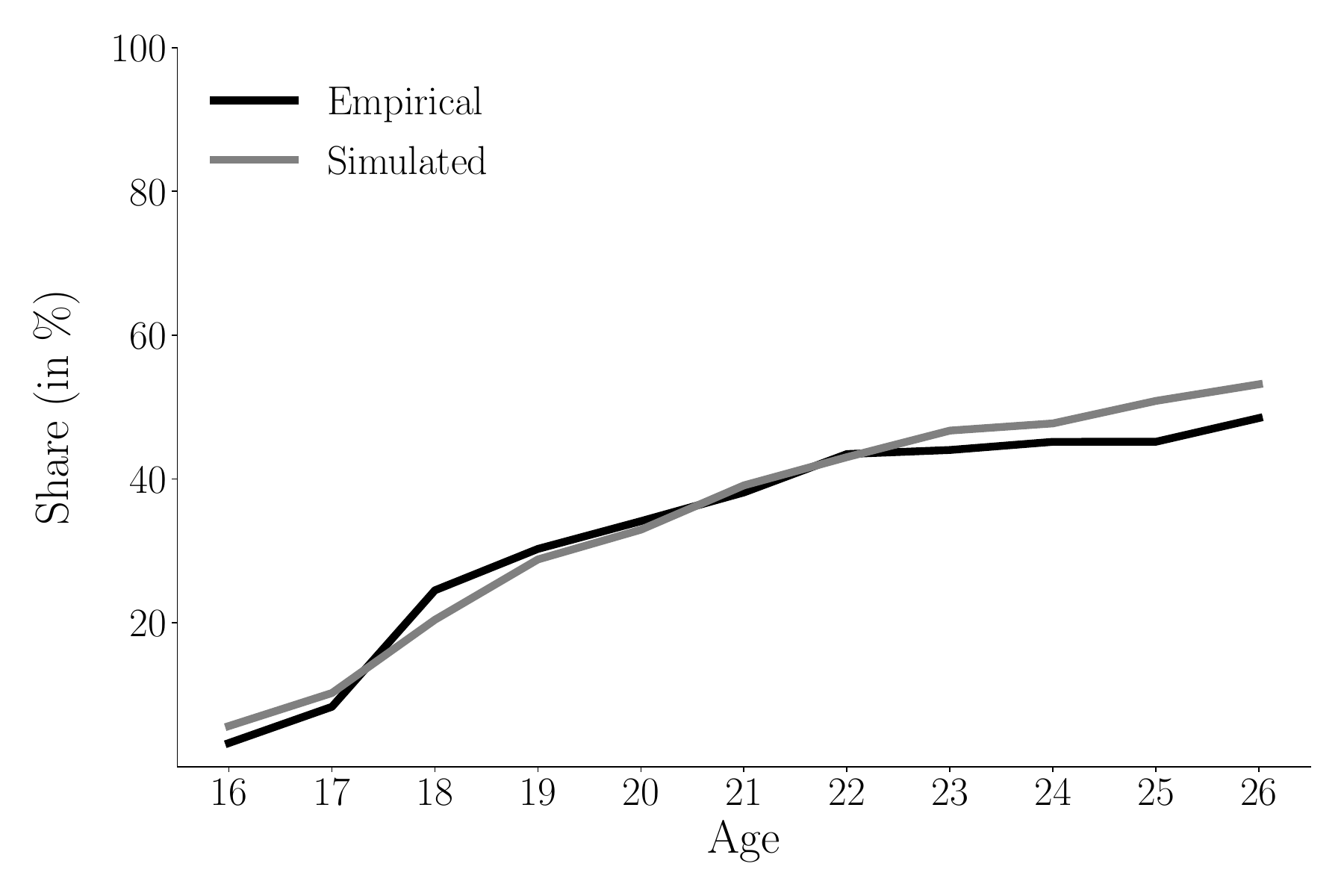}}}
\caption{Model fit}\label{Model fit}
\end{figure}

\noindent Figure \ref{Model fit} shows the overall agreement between the empirical data and a dataset simulated using the estimated model parameters. We show average wages and the share of individuals choosing a blue-collar occupation over time. The results are based on a simulated sample of $10,000$ individuals. Additional model fit statistics are available in the Appendix.
\FloatBarrier

\noindent We adhere to the procedure outlined by the authors of the original paper and use the estimated model to conduct the ex-ante evaluation of a  $\$2,000$ tuition subsidy on educational attainment. We simulate a sample of  $10,000$ individuals using the point estimates and compare completed schooling to a sample of the same size, but with a reduction of $\hat{\beta}_{tc_1}$ by $\$2,000$. The subsidy increases average final schooling by 0.65 years. College graduation increases by 13 percentage points and high school graduation rates improve by 4 percentage points.
%---------------------------------------------------------------------------------------------------
%---------------------------------------------------------------------------------------------------
\subsection{Confidence set bootstrap}
%---------------------------------------------------------------------------------------------------
%---------------------------------------------------------------------------------------------------
The construction of confidence sets for counterfactuals in many structural models poses two distinct challenges. First, the computational burden of even a single estimation of the model is considerable. This makes the application of a standard bootstrap approach \citep{Efron.1979} infeasible. Second, the nonlinear mapping from the parameters of the model to the counterfactual predictions often has kinks or is truncated. For example, in our case, the predicted impact of a tuition subsidy is bounded from below by zero. This violates the smoothness requirements of the delta method.\\

\noindent We use the Confidence Set (CS) bootstrap to construct the confidence set of the counterfactual. Although the CS bootstrap was originally proposed in \citet{Rao.1973}, it has only recently been formalized by \cite{Woutersen.2019}. Its application does not require repeated estimations of the model, as it uses the asymptotic normal distribution of the estimator for $\hat{\btheta}$. Furthermore, its validity does not depend on the differentiability of the prediction function.\footnote{See \citet{Reich.2020} for a critical assessment of confidence sets based on asymptotic arguments. They advocate the use of likelihood-ratio confidence intervals instead and set up their computation as a constraint optimization problem.}\\

\noindent Algorithm \ref{Confidence Set bootstrap} provides a concise description of the steps involved, where $\chi_l^2(1 - \alpha)$ is the quantile function for probability $1 - \alpha$ of the chi-square distribution with $l$ degrees of freedom.\\

\floatname{algorithm}{\sffamily\small Algorithm}
\begin{algorithm}[!th]
	\caption{\small\!\textbf{.\:\:}\textsf{\strut Confidence Set bootstrap}}\label{Confidence Set bootstrap}
	\begin{algorithmic}\vspace{0.3cm}

    \For{$m = 1, \hdots, M$}

    \State Draw $\hat{\btheta}_m \sim \N(\hat{\btheta}, \hat{\bs{\Sigma}})$

    \If{$(\hat{\btheta}_m - \hat{\btheta} )^\prime \hat{\Sigma}^{-1} (\hat{\btheta}_m - \hat{\btheta}) \leq \chi_l^2(1 - \alpha)$}
      \State Compute $\hat{y}_{g,m} = \M_g(\hat{\btheta}_m )$
			\State Add $\hat{y}_{g,m}$ to sample $Y=\{\hat{y}_{g,1}, \hdots, \hat{y}_{g,m-1}\}$
  \EndIf

		\EndFor
    \State Set $\Theta_{y_g}(\alpha) = [\min(Y), \max(Y)]$
		\vspace{0.3cm}\end{algorithmic}
\end{algorithm}\FloatBarrier

\noindent To summarize, we draw a large sample of $M$ parameters from the estimated asymptotic normal distribution of our estimator with mean $\hat{\btheta}$ and covariance matrix $\hat{\bs{\Sigma}}$, accepting only those draws that are elements of the confidence set of the model parameters. We then compute the counterfactual for all remaining draws and calculate the confidence set for the counterfactual based on its lowest and highest value.\\

\noindent The CS bootstrap poses a considerable computational challenge. In many applications, including our own, a single prediction of a counterfactual takes several minutes. At the same time, the number of parameter samples must be large to ensure that the minimum and maximum values for the counterfactual prediction are reliable. However, the algorithm is amenable to parallelization using modern high-performance computational resources by processing each of the $M$ parameter draws independently.\\

\noindent Our uncertainty sets then take the following form:
\begin{align*}\begin{array}{ll}
\U(\alpha) & \equiv\,\,  \left\{\btheta \in\bTheta: (\btheta - \hat{\btheta} )^\prime \hat{\Sigma}^{-1} (\btheta - \hat{\btheta}) \leq \chi_l^2(1 - \alpha)\right\}\\
\U_{y_g}(\alpha)     &\equiv\,\, \left\{M_g(\btheta): (\btheta - \hat{\btheta} )^\prime \hat{\Sigma}^{-1} (\btheta - \hat{\btheta}) \leq \chi_l^2(1 - \alpha), \btheta\in\bTheta\right\}.
\end{array}
\end{align*}

%!TEX root = ../main.tex
%---------------------------------------------------------------------------------------------------
%---------------------------------------------------------------------------------------------------
\FloatBarrier\section{Results}\label{Results}
%---------------------------------------------------------------------------------------------------
%---------------------------------------------------------------------------------------------------
Turning to the presentation of our results, we focus on the impact of a $\$2,000$ tuition subsidy on completed schooling and use the 90\% uncertainty set to measure the degree of uncertainty.  All our results potentially depend on the size of the uncertainty set. In practice, policy-makers choose the uncertainty set's size in line with their underlying preferences - the more desirable protection against unfavorable outcomes is, the larger the uncertainty set will be.\footnote{In a different setting, \citet{Blesch.2021} conduct an ex-ante performance evaluation  of the statistical decision functions over the whole parameter space \citep{Wald.1950,Manski.2021}.}\\

\noindent All results are based on $30,000$ draws from the asymptotic normal distribution of our parameter estimates. We follow \citet{Keane.1997} and start by analyzing the prediction for a general subsidy. Then we turn to the situation where we use endowment types for policy targeting. Throughout our analysis, we postulate a linear utility function for the policy-maker.
%---------------------------------------------------------------------------------------------------
\subsection{General subsidy}
\begin{figure}[ht!]\centering
\scalebox{0.35}{\includegraphics{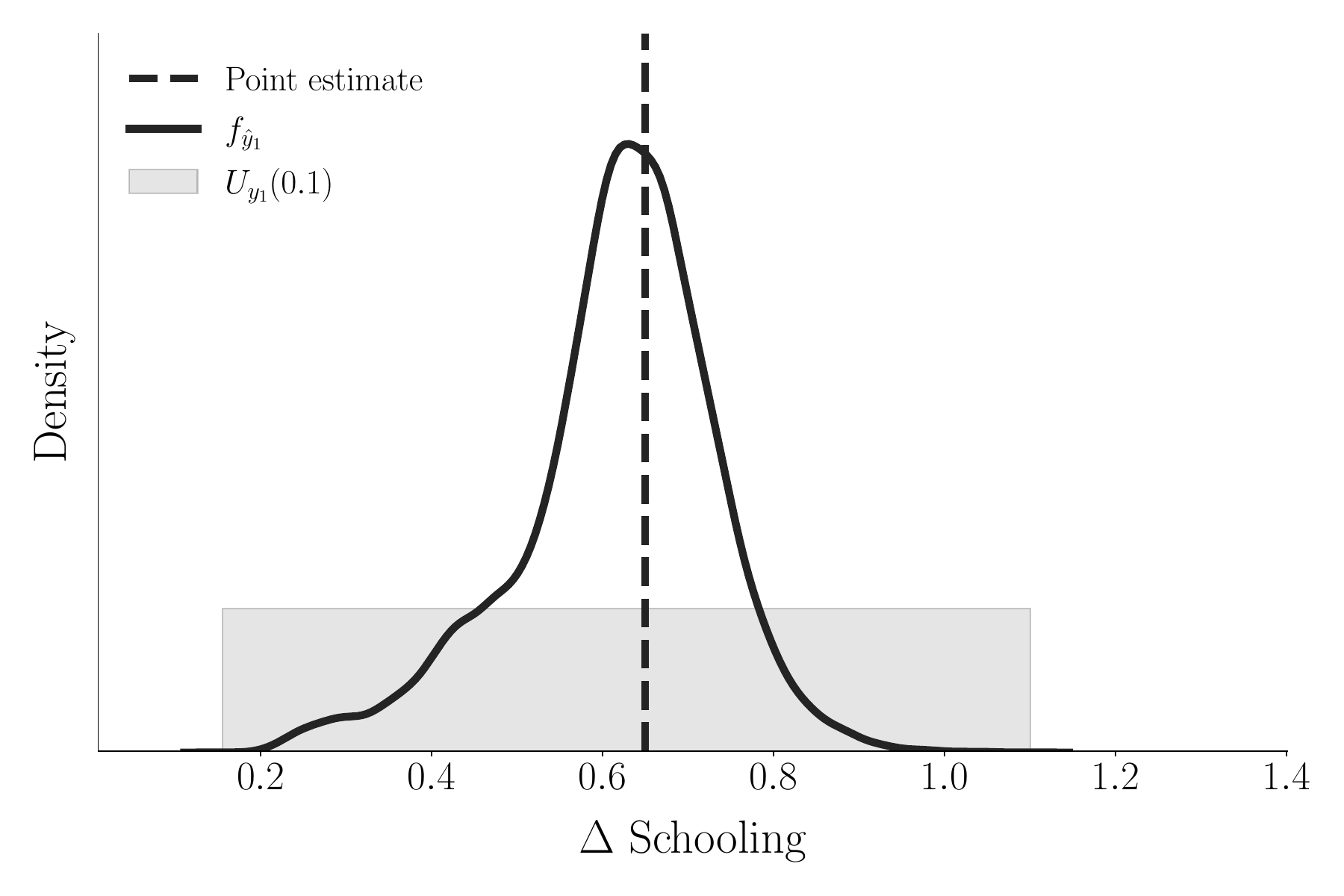}}
\caption{General subsidy}\label{General subsidy}
\end{figure}

%---------------------------------------------------------------------------------------------------
\noindent  Figure \ref{General subsidy} explores the impact prediction for a general tuition subsidy. We show the point prediction, its sampling distribution, and the uncertainty set. At the point estimate, average schooling increases by $0.65$ years. However, there is considerable uncertainty about the prediction, as the uncertainty set ranges from $0.15$ to $1.10$ years.\\

\begin{figure}[ht!]\centering
\scalebox{0.35}{\includegraphics{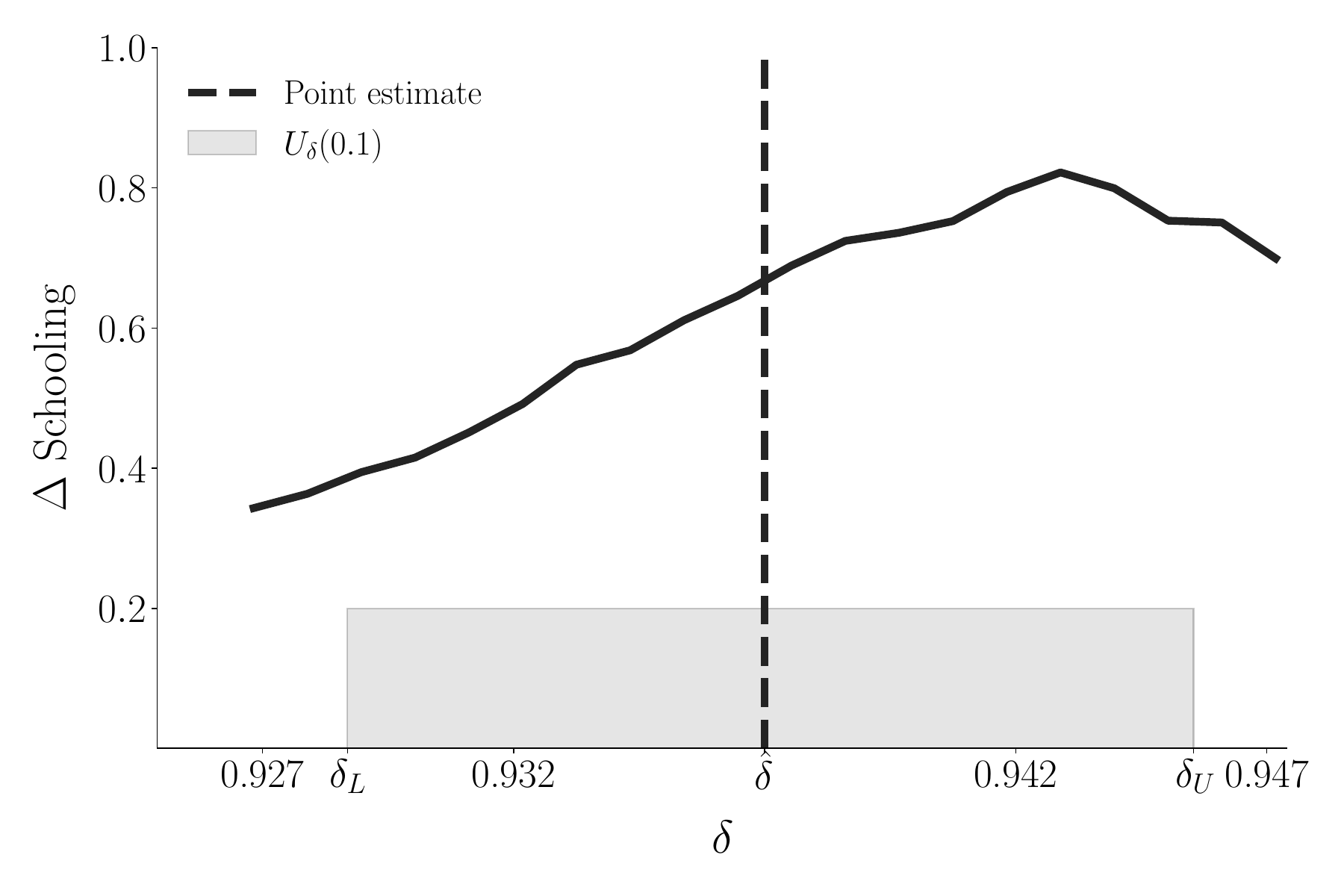}}
\caption{Time preference}\label{Time preference}
\end{figure}

\FloatBarrier
\noindent In Figure \ref{Time preference}, we trace the effect of the discount rate $\delta$ on the subsidy's impact over the uncertainty set, while keeping all other parameters at their point estimate. Initially, as  $\delta$ increases, so does the policy's impact as individuals value the long-term benefits from increasing their level of schooling more and more. However, for high levels of the discount factor, the policy's impact starts to decrease as most individuals already complete a high school or college degree even without the subsidy.
\FloatBarrier
%---------------------------------------------------------------------------------------------------
\subsection{Targeted subsidy}
%---------------------------------------------------------------------------------------------------
So far, we restricted the analysis to a general subsidy available to the whole population and the average predicted impact. We now examine the setting in which a policy-maker can target individuals based on the type of their initial endowment. The importance of early endowment heterogeneity in shaping economic outcomes over the life-cycle is the most important finding from \citet{Keane.1997}. It served as motivation for a host of subsequent research on the determinants of skill heterogeneity among adolescents \citep{Caucutt.2020,Erosa.2010,Todd.2007}.\\

\noindent To ease the exposition, we initially focus our discussion of results on Type 1 and Type 3 individuals. We later rank policies targeting either of the four types based on the different decision-theoretic criteria. Additional results are available in our Appendix.\\

\begin{figure}[h!]\centering
\subfloat[Life-cycle choices]{\scalebox{0.25}{\includegraphics{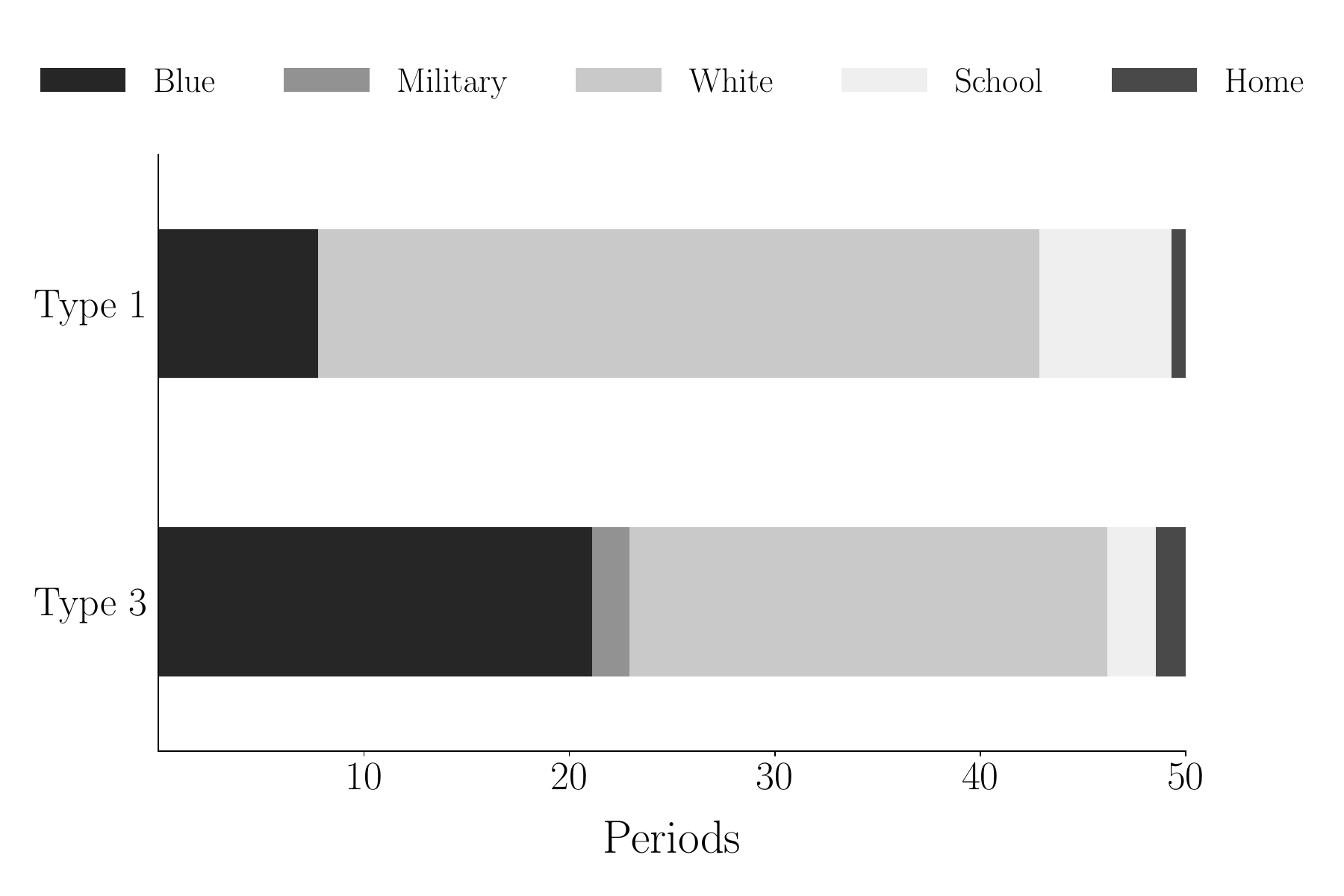}}\label{Type choices}}
\subfloat[Completed schooling]{\scalebox{0.25}{\includegraphics{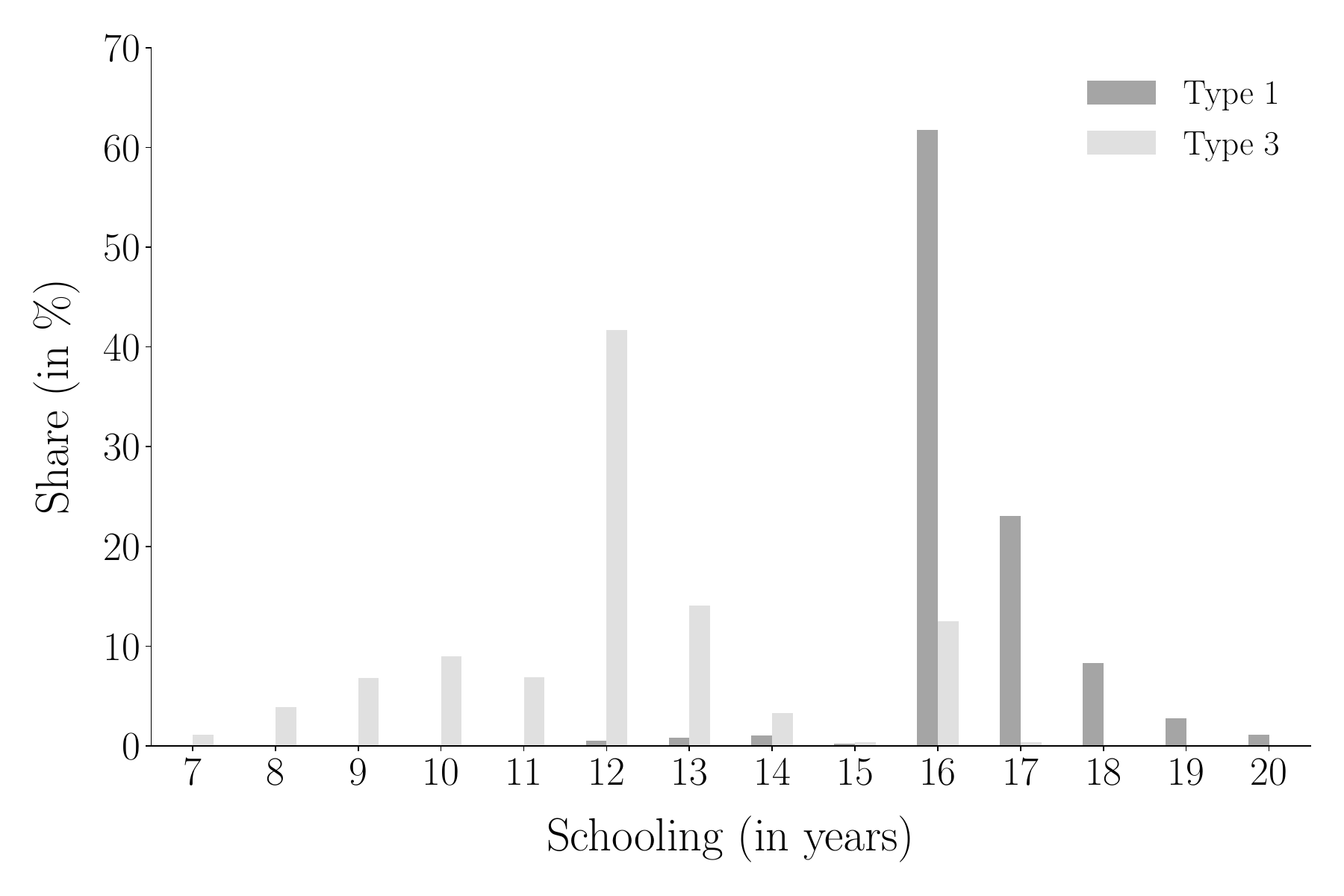}}\label{Type schooling}}\\
\caption{Type heterogeneity}\label{Type heterogeneity}
\end{figure}

\noindent Figure \ref{Type heterogeneity} confirms that life-cycle choices differ considerably by initial endowment type. On the left, we show the number of periods the two types spend on average in each of the five alternatives. Those characterized as Type 1 individuals spend more than six years on their education even after entering the model. Type 3 individuals, on the other hand, extend their academic pursuits for only an additional two years. This difference translates into very different labor market experiences. While Type 1 individuals work for about 35 years in a white-collar occupation, Type 3 workers switch more frequently between white and blue-collar occupations and spend a comparable amount of time working in either occupation -- approximately 44 years split equally among white and blue-collar occupations. Both types only spend a short time at home.
\FloatBarrier
\noindent On the right, we show the distribution of final schooling for both types. Years of schooling are considerably higher for Type 1 individuals with an average of more than 16 years compared to only 12 years for those identified as Type 3 individuals. Nearly all Type 1 individuals enroll in college and most graduate with a degree.\\

\begin{figure}[h!]\centering
\subfloat[Type 1]{\scalebox{0.25}{\includegraphics{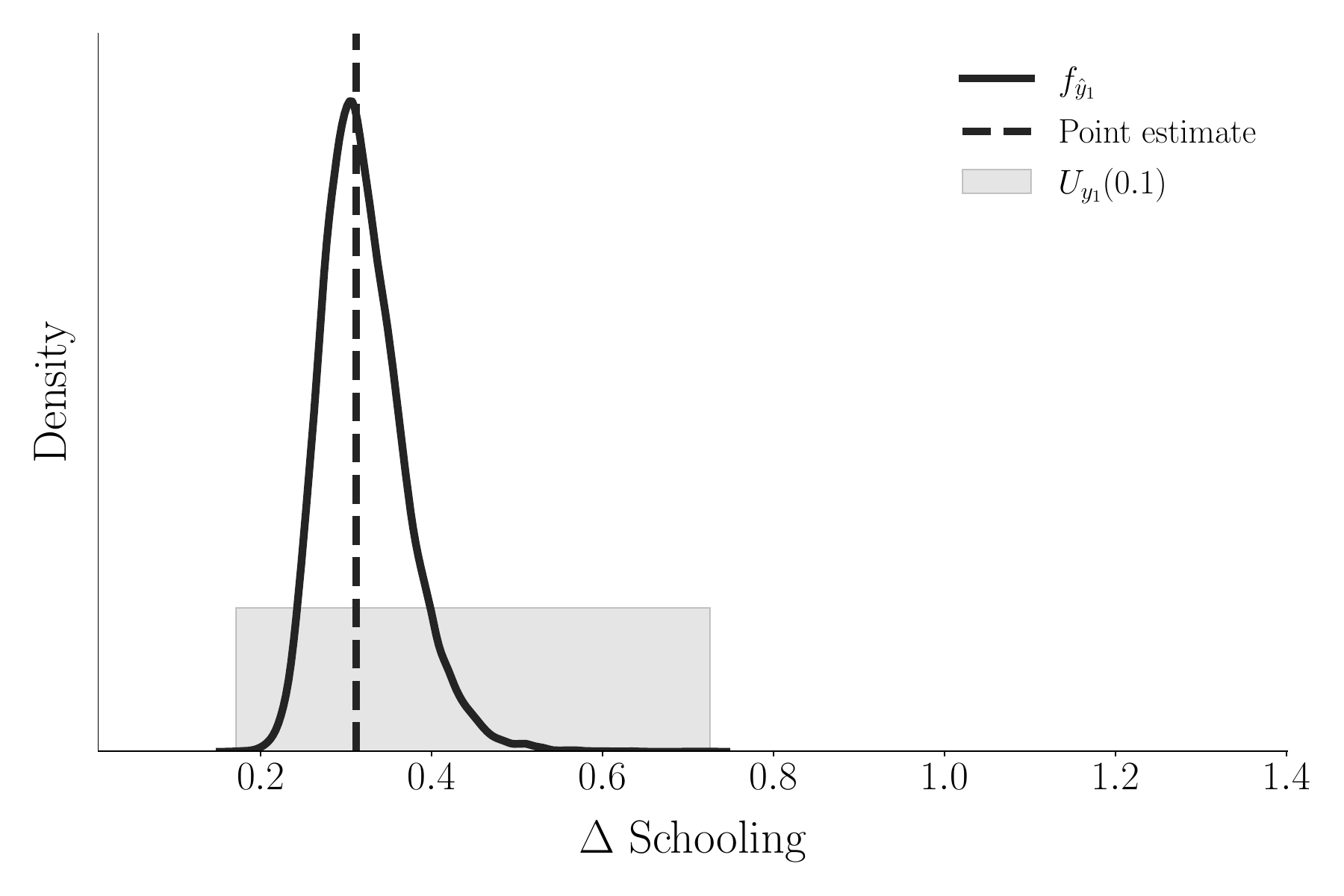}}\label{type 0}}
\subfloat[Type 3]{\scalebox{0.25}{\includegraphics{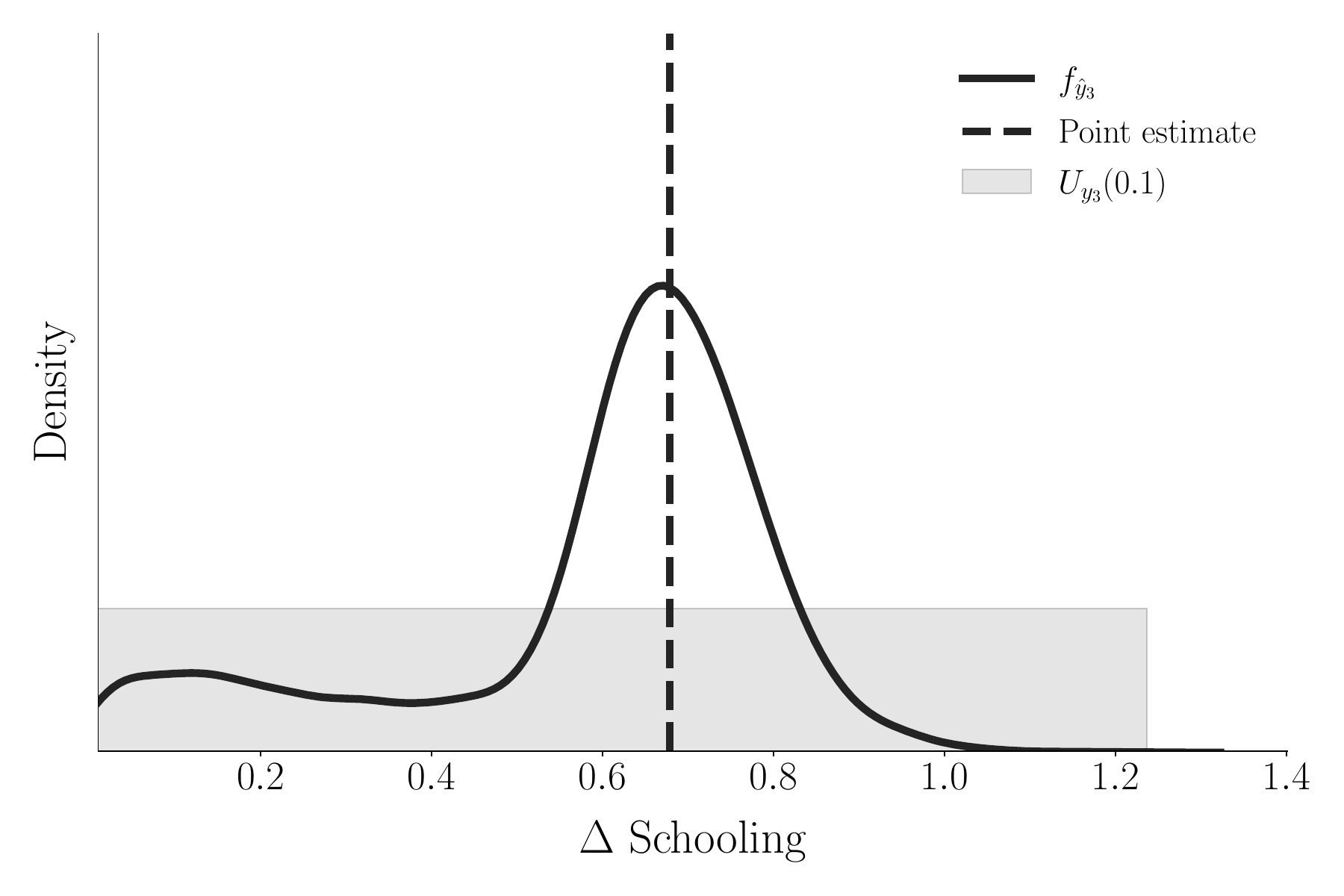}}\label{type 2}}\\
\caption{Targeted subsidy}\label{Targeted subsidy}
\end{figure}

\noindent Figure \ref{Targeted subsidy} provides a visualization of our core results for a targeted subsidy. At the point estimates, the predicted impact is considerably lower for Type 1 than Type 3. However, the prediction uncertainty is much larger for Type 3 compared to Type 1. The uncertainty set for Type 3 ranges all the way from $0$ to $1.2$ years, while the prediction for Type 1 is between $0.18$ and $0.75$.\\
\FloatBarrier
\noindent This heterogeneity in impact and prediction uncertainty follows directly from the underlying economics of the model. Type 1 individuals are already more likely to have a college degree before the subsidy, and thus, the predicted impact is smaller. Alternatively, Type 1 individuals affected by the subsidy are in the middle of pursuing a college education and thus directly benefit from it. Since Type 3 individuals are at the lower end of the schooling distribution, a tuition subsidy can considerably increase their level of schooling. Whether the subsidy succeeds in doing so, however, remains uncertain.\\

\noindent We now consider the policy option to target Type 2 and Type 4 as well. Their point predictions are actually highest with an additional $0.81$ years on average for Type 2 and $0.75$ years for Type 4. However, both predictions are fraught with uncertainty. For Type 2 the uncertainty set ranges from $0.17$ to $1.3$, while for Type 4 it starts at zero and spans all the way to $1.18$.\\

\begin{figure}[h!]\centering
\scalebox{0.35}{\includegraphics{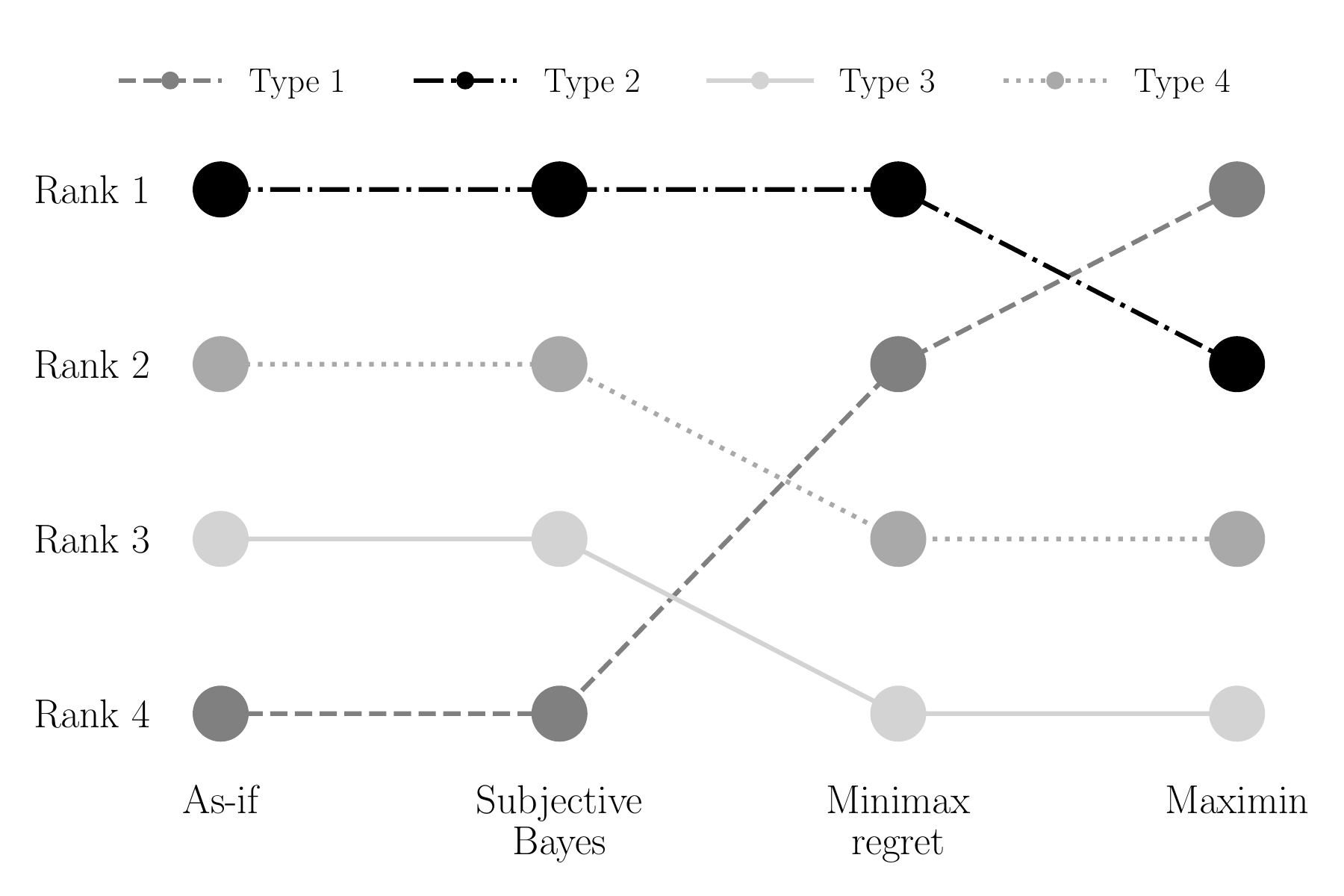}}
\caption{Policy ranking}\label{Ranking}
\end{figure}

\noindent Figure \ref{Ranking} shows the policy alternative's ranking by the decision-theoretic criteria we discussed in Section \ref{Statistical decision theory}. Ranking alternatives using as-if optimization is straightforward. A policy targeting Type 2 is the most preferred alternative, while a focus on Type 1 is the least attractive. However, once we account for the presence of uncertainty in the predictions, a more nuanced picture emerges. Moving from as-if optimization to a subjective Bayes criterion using a uniform distribution over the uncertainty set does not change the ordering. However, once a decision-maker is concerned with performance across the whole range of values in the uncertainty set -- we move to the minimax regret or maximin criterion -- a policy targeting Type 1 becomes more and more attractive despite its low point prediction because its worst-case utility is highest.\\
\FloatBarrier
\noindent In general, framing policy advice as a decision problem under uncertainty shows that there are many different ways of making reasonable decisions. The ranking of policies varies depending on the decision criteria. Not only that, but due to the necessary ex-post nature of our implementation, the ranking for a given criteria also depends on the choice of $\alpha$. The selection of $\alpha$ is part of the decision problem: the more a policy-maker is concerned about worst-case scenarios, the smaller the appropriate value for $\alpha$ will be. After deciding on a preferred decision rule, we suggest performing a sensitivity analysis around the selected $\alpha$ value by checking how much the policy ranking varies within a neighborhood.

% !TEX root = ../main.tex
%---------------------------------------------------------------------------------------------------
\FloatBarrier\section{Conclusion}\label{Conclusion}
%---------------------------------------------------------------------------------------------------
We develop a generic approach that addresses parametric uncertainty when using models to inform policy-making. We propose a decision-theoretic analysis of computationally demanding structural models based on uncertainty sets. We construct the uncertainty sets from empirical estimates and ensure their computational tractability by using the confidence set bootstrap. We revisit the seminal work by \citet{Keane.1997} to document the empirical relevance of prediction uncertainty and showcase our analysis. Focusing on their ex-ante evaluation of a tuition subsidy, we report considerable uncertainty in the policy's impact on completed schooling. We show how a policy-maker's preferred policy depends on the choice of alternative formal rules for decision-making under uncertainty.\\

\noindent In our ongoing research, we pursue three avenues for further improvements. First, we link our work with the literature on inference under (local) model misspecification to refine the construction of our uncertainty sets. For example, \citet{Armstrong.2021} and \citet{Bonhomme.2020} propose different methods for taking misspecification into account when constructing confidence sets. Second, we incorporate ideas from the literature on global sensitivity analysis \citep{Razavi.2021} to identify the parameters most responsible for uncertainty in predictions. The attribution of importance based on Shapely values, familiar to economists from game theory, appears promising \citep{Owen.2014, Shapley.1953} as well. Third, we address our analysis's computational burden using surrogate modeling \citep{Forrester.2008}, which emulates the full model's behavior at a negligible cost per run and allows us to determine prediction uncertainty using a nonparametric bootstrap procedure.

%\bibliographystyle{apalike}
%\bibliography{literature}

%!TEX root = ../main.tex
\begin{appendices}\setcounter{page}{1}
\renewcommand{\thepage}{\thesection-\arabic{page}}
%---------------------------------------------------------------------------------------------------
%---------------------------------------------------------------------------------------------------
\FloatBarrier\section{Appendix}\label{Appendix}
%---------------------------------------------------------------------------------------------------
%---------------------------------------------------------------------------------------------------
The Appendix contains details on our computational implementation, the estimation dataset, and additional results.

%!TEX root = ../main.tex
%---------------------------------------------------------------------------------------------------
\subsection{Computation}\label{Appendix computation}\FloatBarrier
%---------------------------------------------------------------------------------------------------
Using the same computational implementation as \citetAppndx{Keane.1997}, we outline the immediate utility functions for each of the five alternatives. We first focus on their common structure and then present their parameterization. We also provide the economic motivation for their specification.\\

\noindent We follow individuals over their working life from age 16 until retirement at age 65. Each decision period $t = 16, \dots, 65$  represents a school year. Individuals can select one of five alternatives from the set of admissible actions $a\in\mathcal{A}$. They can decide to either work in a blue-collar or white-collar occupation ($a = 1, 2$), serve in the military $(a = 3)$, attend school $(a = 4)$, or stay at home $(a = 5)$.\\

\noindent Individuals differ with respect to their initial level of completed schooling $h_{16}$, and they possess one of four $\mathcal{J} = \{1, \hdots, 4\}$ alternative-specific skill endowments $\bm{e} = \left(e_{j,a}\right)_{\mathcal{J} \times \mathcal{A}}$.\\

\noindent The immediate utility $u_a(\cdot)$ of each alternative consists of a non-pecuniary utility $\zeta_a(\cdot)$ and, at least for the working alternatives, an additional wage component $w_a(\cdot)$, both of which depend on the level of human capital as measured by their occupation-specific work experience $\bm{k}_t = \left(k_{a,t}\right)_{a\in\{1, 2, 3\}}$, years of completed schooling $h_t$, and alternative-specific skill endowment $\bm{e}$. The immediate utility functions are influenced by last-period choices $a_{t -1}$ and alternative-specific productivity shocks $\bm{\epsilon}_t = \left(\epsilon_{a,t}\right)_{a\in\mathcal{A}}$ as well. Their general form is given by:
\begin{align*}
u_a(\cdot) =
\begin{cases}
    \zeta_a(\bm{k}_t, h_t, t, a_{t -1})  + w_a(\bm{k}_t, h_t, t, a_{t -1}, e_{j, a}, \epsilon_{a,t})                & \text{if}\, a \in \{1, 2, 3\}  \\
    \zeta_a(\bm{k}_t, h_t, t, a_{t-1}, e_{j,a}, \epsilon_{a,t})                                                  &  \text{if}\, a \in \{4, 5\}.
\end{cases}
\end{align*}
Work experience $\bm{k}_t$  and years of completed schooling $h_t$ evolve deterministically:
\begin{align*}
	k_{a,t+1} & = k_{a,t} + \ind[a_t = a]  \qquad \text{if}\, a \in \{1, 2, 3\} \\
	h_{t + 1\phantom{,a}} & = h_{t\phantom{,a}} +   \ind[a_t = 4].
\end{align*}
\noindent The productivity shocks are uncorrelated across time and follow a multivariate normal distribution with mean $\bm{0}$ and covariance matrix $\bm{\Sigma}$. Given the structure of the utility functions and the distribution of the shocks, the state at time $t$ is $s_t = \{\bm{k}_t, h_t, t, a_{t -1}, \bm{e},\bm{\epsilon}_t\}$.\\

\noindent Empirical and theoretical research from specialized disciplines within economics informs the exact specification of $u_a(\cdot)$. We now discuss each of its components in detail.
%-------------------------------------------------------------------------------
\subsubsection*{Non-pecuniary utility}
%-------------------------------------------------------------------------------
We begin by presenting the parameterization of the non-pecuniary utility for all five alternatives.
%-------------------------------------------------------------------------------
\paragraph{Blue-collar}
%-------------------------------------------------------------------------------
Equation (\ref{eq:NonWageBLueCollar}) shows the parameterization of the non-pecuniary utility from working in a blue-collar occupation:
\begin{align}\label{eq:NonWageBLueCollar}
\zeta_{1}(\bm{k}_t, h_t, a_{t-1})  = \alpha_1  &+ c_{1,1} \cdot \ind[a_{t-1} \neq 1] + c_{1,2} \cdot \ind[k_{1,t} = 0] \\ \nonumber
                            & + \vartheta_1 \cdot \ind[h_t \geq 12] + \vartheta_2 \cdot \ind[h_t \geq 16] + \vartheta_3 \cdot \ind[k_{3,t} = 1].
\end{align}
A constant $\alpha_1$ captures the net monetary equivalent of on-the-job amenities. Non-pecuniary utility includes mobility and search costs $c_{1,1}$, which are higher for individuals who had previously never worked in a blue-collar occupation, $c_{1,2}$, and captures returns from high school, $\vartheta_1$, and college degrees, $\vartheta_2$. Additionally, there is a detrimental effect of prematurely leaving the military after one year, $\vartheta_3$.
%-------------------------------------------------------------------------------
\paragraph{White-collar}
%-------------------------------------------------------------------------------
The non-pecuniary utility from working in a white-collar occupation is specified analogously. Equation (\ref{eq:UtilityWhiteCollar}) shows its parameterization:
\begin{align}\label{eq:UtilityWhiteCollar}
\zeta_{2}( \bm{k}_t, h_t, a_{t-1} ) = \,\alpha_2 & + c_{2,1} \cdot \ind[a_{t-1} \neq 2] + c_{2,2} \cdot \ind[k_{2,t} = 0]\\\nonumber
                            & + \vartheta_1 \cdot \ind[h_t \geq 12] + \vartheta_2 \cdot \ind[h_t \geq 16] + \vartheta_3 \cdot \ind[k_{3,t} = 1].
\end{align}
%-------------------------------------------------------------------------------
\paragraph{Military}
%-------------------------------------------------------------------------------
Equation (\ref{eq:UtilityMilitary}) shows the parameterization of the non-pecuniary utility from working in the military:
\begin{align}\label{eq:UtilityMilitary}
\zeta_{3}( k_{3.t}, h_t)  = \,& c_{3,2} \cdot \ind[k_{3,t} = 0]+ \vartheta_1 \cdot \ind[h_t \geq 12] + \vartheta_2 \cdot \ind[h_t \geq 16].
\end{align}
Although search costs $c_{3, 1} = 0$ are absent, there is a mobility cost if an individual has never previously served in the military, $c_{3,2}$. Individuals still experience a non-pecuniary utility from completing high school, $\vartheta_1$, and college, $\vartheta_2$.
%-------------------------------------------------------------------------------
\paragraph{School}
%-------------------------------------------------------------------------------
Equation (\ref{eq:UtilitySchooling}) shows the parameterization of the non-pecuniary utility from schooling:
\begin{align}\label{eq:UtilitySchooling}
	\zeta_4(k_{3,t}, h_t, t, a_{t-1}, e_{j,4}, \epsilon_{4,t})  = e_{j,4} & + \beta_{tc_1} \cdot \ind[h_t \geq 12] + \beta_{tc_2} \cdot \ind[h_t \geq 16]   \\\nonumber
    							  & + \beta_{rc_1} \cdot \ind[a_{t-1} \neq 4, h_t < 12]  \\\nonumber
    							  & + \beta_{rc_2} \cdot \ind[a_{t-1} \neq 4, h_t \geq 12]+ \gamma_{4,4} \cdot t  \\\nonumber
     							  & + \gamma_{4,5} \cdot \ind[t < 18] + \vartheta_1 \cdot \ind[h_t \geq 12]  \\\nonumber
      							& + \vartheta_2 \cdot \ind[h_t \geq 16]+ \vartheta_3 \cdot \ind[k_{3,t} = 1] + \epsilon_{4,t}.
\end{align}
There are direct costs for pursuing higher education, which primarily take the form of college, $\beta_{tc_1}$, and graduate school tuition fees, $\beta_{tc_2}$. The decision to leave school is reversible, but entails adjustment costs that differ by schooling category ($\beta_{rc_1}, \beta_{rc_2}$). Education is defined by time spent in school, not by formal credentials acquired. Once individuals reach a certain amount of schooling, they acquire a degree. There is no uncertainty about grade completion \citepAppndx{Altonji.1993} and no part-time enrollment. Individuals value the completion of high school and college ($\vartheta_1, \vartheta_2$).
%-------------------------------------------------------------------------------
\paragraph{Home}
%-------------------------------------------------------------------------------
Equation (\ref{eq:UtilityHome}) shows the parameterization of the non-pecuniary utility from staying at home:
\begin{align}\label{eq:UtilityHome}
	\zeta_5(k_{3,t}, h_t, t, e_{j,5}, \epsilon_{5,1}) =  e_{j,5} & + \gamma_{5,4} \cdot \ind[18 \leq t \leq 20] + \gamma_{5,5} \cdot \ind[t \geq 21] \\ \nonumber
    							   & +\vartheta_{1} \cdot \ind[h_t \geq 12] + \vartheta_{2} \cdot \ind[h_t \geq 16] \\ \nonumber
    							   & +  \vartheta_3 \cdot \ind[k_{3,t} = 1]+ \epsilon_{5,t}.
\end{align}
Staying at home as a young adult, $\gamma_{5, 4}$, is less stigmatized than doing so as an older individual, $\gamma_{5,5}$. Possessing a degree $(\vartheta_1, \vartheta_2)$ or leaving the military prematurely, $\vartheta_3$, influences the immediate utility as well.
%-------------------------------------------------------------------------------
\subsubsection*{Wage component}
%-------------------------------------------------------------------------------
The wage component $w_{a}(\cdot)$ for the working alternatives is given by the product of the market-equilibrium rental price $r_{a}$ and an occupation-specific skill level $x_{a}(\cdot)$. The latter is determined by the overall level of human capital:
\begin{align*}
w_{a}(\cdot) = r_{a} \, x_{a}(\cdot).
\end{align*}
This specification leads to a standard logarithmic wage equation in which the constant term is the skill rental price $\ln(r_{a})$ and wages follow a log-normal distribution.\\

\noindent The occupation-specific skill level $x_{a}(\cdot)$ is determined by a skill production function, which includes a deterministic component $\Gamma_a(\cdot)$ and a multiplicative stochastic productivity shock $\epsilon_{a,t}$:
\begin{align}
    x_{a}(\bm{k}_t, h_t, t, a_{t-1}, e_{j, a}, \epsilon_{a,t}) & = \exp \big( \Gamma_{a}(\bm{k}_t,  h_t, t, a_{t-1}, e_{j,a}) \cdot \epsilon_{a,t} \big). \nonumber
\end{align}
%-------------------------------------------------------------------------------
\paragraph{Blue-collar}
%-------------------------------------------------------------------------------
Equation (\ref{eq:SkillLevelBlueCollar}) shows the parameterization of the deterministic component of the skill production function:
\begin{align}\label{eq:SkillLevelBlueCollar}
    \Gamma_1(\bm{k}_t, h_t, t, a_{t-1}, e_{j, 1}) = e_{j,1} & + \beta_{1,1} \cdot h_t + \beta_{1, 2} \cdot \ind[h_t \geq 12] \\ \nonumber
                                  & + \beta_{1,3} \cdot \ind[h_t\geq 16] + \gamma_{1, 1} \cdot  k_{1,t} + \gamma_{1,2} \cdot  (k_{1,t})^2 \\ \nonumber
                                & + \gamma_{1,3} \cdot  \ind[k_{1,t} > 0] + \gamma_{1,4} \cdot  t + \gamma_{1,5} \cdot \ind[t < 18]\\ \nonumber
                                  & + \gamma_{1,6} \cdot \ind[a_{t-1} = 1] + \gamma_{1,7} \cdot  k_{2,t} + \gamma_{1,8} \cdot  k_{3,t}. \nonumber
\end{align}

\noindent There are several notable features. The first part of the skill production function is motivated by \citetAppndx{Mincer.1974} and, hence, linear in years of completed schooling, $\beta_{1,1}$, quadratic in experience ($\gamma_{1,1}, \gamma_{1,2}$), and separable between the two of them. There are so-called sheep-skin effects \citepAppndx{Hungerford.1987, Jaeger.1996} associated with completing a high school, $\beta_{1,2}$, or graduate education, $\beta_{1,3}$, which capture the impact of completing a degree beyond the associated years of schooling. There is also a first-year blue-collar experience effect $\gamma_{1,3}$. Additionally, job skills depreciate for blue-collar workers, who were unemployed in the previous period, $\gamma_{1,6}$. All forms of work experience ($\gamma_{1,7}, \gamma_{1,8}$) are transferable.
%-------------------------------------------------------------------------------
\paragraph{White-collar}
%-------------------------------------------------------------------------------
The wage component from working in a white-collar occupation is specified analogously. Equation (\ref{eq:SkillLevelWhiteCollar}) shows the parameterization of the deterministic component of the skill production function:
\begin{align}\label{eq:SkillLevelWhiteCollar}
    \Gamma_2(\bm{k}_t, h_t, t, a_{t-1}, e_{j,2}) = e_{j,2} & + \beta_{2,1} \cdot h_t + \beta_{2, 2} \cdot \ind[h_t \geq 12]  \\\nonumber
    							 & + \beta_{2,3} \cdot \ind[h_t\geq 16]+ \gamma_{2, 1} \cdot  k_{2,t} + \gamma_{2,2} \cdot  (k_{2,t})^2  \\\nonumber
                                   & + \gamma_{2,3} \cdot  \ind[k_{2,t} > 0]+ \gamma_{2,4} \cdot  t + \gamma_{2,5} \cdot \ind[t < 18] \\\nonumber
                                  & + \gamma_{2,6} \cdot  \ind[a_{t-1} = 2]  + \gamma_{2,7} \cdot  k_{1,t} + \gamma_{2,8} \cdot  k_{3,t}.
\end{align}
%-------------------------------------------------------------------------------
\paragraph{Military}
%-------------------------------------------------------------------------------
Equation (\ref{eq:SkillLevelMilitary}) shows the parameterization of the deterministic component of the skill production function:
\begin{align}\label{eq:SkillLevelMilitary}
    \Gamma_3( k_{3,t}, h_t, t, e_{j,3}) = e_{j,3} & + \beta_{3,1} \cdot h_t  \\\nonumber
	               \nonumber &+ \gamma_{3,1} \cdot  k_{3,t} + \gamma_{3,2} \cdot (k_{3,t})^2 + \gamma_{3,3} \cdot \ind[k_{3,t} > 0]\\\nonumber
									 & + \gamma_{3,4} \cdot t + \gamma_{3,5} \cdot \ind[t < 18].
\end{align}
Unlike the civilian sector, there are no sheep-skin effects from completing military training, ($\beta_{3,2} = \beta_{3,3}= 0$). Furthermore, the previous occupational choice has no influence ($\gamma_{3,6}= 0$), and any experience other than military is non-transferable ($\gamma_{3,7} = \gamma_{3,8} = 0$).

\begin{Remark} Our parameterization for the immediate utility of serving in the military differs from \citetAppndx{Keane.1997}, as we remain unsure about their exact specification. The authors state in Footnote 31 (p.\,498) that the constant for the non-pecuniary utility $\alpha_{3,t}$ depends on age. However, we are unable to determine the precise nature of the relationship. Equation (C3) (p.\,521) also indicates no productivity shock $\epsilon_{a,t}$ in the wage component. Table~7 (p.\,500) reports such estimates.
\end{Remark}

\noindent Table \ref{Model parameters} presents an overview of the model's parameters.

\begin{ThreePartTable}
	% Information available at https://ftp.agdsn.de/pub/mirrors/latex/dante/macros/latex/contrib/threeparttablex/threeparttablex.pdf

	\begin{TableNotes}
		\item \textbf{Note:} The above list is an overview of the model parameters. The immediate utilities for the alternatives do not necessarily include all of them.
	\end{TableNotes}
	\begin{longtable}{@{}cll@{}}
		\caption{Overview of parameters in the \citet{Keane.1997} extended model.}
		\label{Model parameters}

		\setlength\extrarowheight{2.5pt}

		% Settings longtable
		\\
		\toprule
		\textbf{Parameter}            &  &  \multicolumn{1}{l}{\textbf{Description}}              \\ \midrule
		\endfirsthead

		% Alternative 1 header for the beginning on next page
		%\midrule
		%Parameter            &  &  \multicolumn{1}{l}{Description}     \\ \midrule

		% Alternative 2 header for the beginning on next page
		\multicolumn{3}{c}{\tikz\draw [thick,dash dot] (0,0) -- (11.5,0);} \\
		\multicolumn{3}{c}{continued on next page }  \vspace{-4pt}\\
		\multicolumn{3}{c}{\tikz\draw [thick,dash dot] (0,0) -- (11.5,0);} \\
		\endfoot

		\multicolumn{3}{c}{\tikz\draw [thick,dash dot] (0,0) -- (11.5,0);} \\
		\multicolumn{3}{c}{continued from previous page} \vspace{-4pt} \\
		\multicolumn{3}{c}{\tikz\draw [thick,dash dot] (0,0) -- (11.5,0);} \\
		\endhead

		\bottomrule
		\insertTableNotes
		\endlastfoot

		% Start table
		\midrule
		\multicolumn{3}{l}{Preference and type-specific parameters}																		 \\ \midrule
		$\delta$ 				&  & discount factor																			  \\
		$e_{j, a}$			&  & initial endowment of type $j$ in alternative $a$ specific skills 	  \\ [7.5pt] \midrule
		\multicolumn{3}{l}{Common parameters immediate utility}												\\ \midrule
		$\alpha_a$           &  & return on non-wage working conditions		   \\
		$\vartheta_1$        &  & non-pecuniary premium for finishing high school                 								    \\
		$\vartheta_2$        &  & non-pecuniary premium for finishing college															    \\
		$\vartheta_3$        &  & non-pecuniary premium for leaving the military early						  \\[7.5pt] \midrule
		\multicolumn{3}{l}{Schooling-related parameters}															   \\ \midrule
		% Work
		$\beta_{a,1}$        &  & return on each additional year of completed schooling 								\\
		$\beta_{a,2}$        &  & skill premium for high school graduates										      \\
		$\beta_{a,3}$        &  & skill premium for college graduates													   	\\
		% Military
		% School
		$\beta_{tc_1}$       &  & tuition costs for high school                      											\\
		$\beta_{tc_2}$       &  & tuition costs for college                          												\\
		$\beta_{rc_1}$       &  & re-entry costs for high school                     										   \\
		$\beta_{rc_2}$       &  & re-entry costs for college                        												   \\
		% Home
		$\beta_{5,2}$        &  & skill premium for high school graduates            									\\
		$\beta_{5,3}$        &  & skill premium for college graduates                									     \\ [7.5pt] \midrule
		\multicolumn{3}{l}{Experience-related parameters}           													 \\
		\midrule
		$\gamma_{a,1}$       &  & return on same-sector experience                 									 \\
		$\gamma_{a,2}$       &  & return squared on same-sector experience         								\\
		$\gamma_{a,3}$       &  & premium for having previously worked in sector        							   \\
		$\gamma_{a,4}$       &  & return on age effect                             											     \\
		$\gamma_{a,5}$       &  & return on age effect for minors               										\\
		$\gamma_{a,6}$       &  & premium for remaining in same sector              								   \\
		$\gamma_{a,7}$       &  & return on civilian cross-sector experience       								    \\
		$\gamma_{a,8}$       &  & return on non-civilian sector experience       										 \\
		$\gamma_{3,1}$       &  & return on same-sector experience                 									  \\
		$\gamma_{3,2}$       &  & return squared on same-sector experience    										 \\
		$\gamma_{3,3}$       &  & premium for having previously worked in sector   										\\
		$\gamma_{3,4}$       &  & return on age effect                             												 \\
		$\gamma_{3,5}$       &  & return on age effect for minors              	   										\\
		$\gamma_{4,4}$       &  & return on age effect                             												 \\
		$\gamma_{4,5}$       &  & return on age effect for minors                  										\\
		$\gamma_{5,4}$       &  & return on age, between 17 and 21                 	  									   \\
		$\gamma_{5,5}$       &  & return on age, older than 21							   										\\[7.5pt] \midrule
		\multicolumn{3}{l}{Mobility and search parameters}          													  \\ \midrule
		$c_{a,1}$            &  & premium for switching to occupation $a$           									   \\
		$c_{a,2}$            &  & premium for working in occupation $a$ for the first time         										  \\
		$c_{3,2}$            &  & premium for serving in the military for the first time								                                            	  \\[7.5pt] \midrule
		\multicolumn{3}{l}{Error correlation}          													  									\\ \midrule
		$\sigma_{a,a}$	&	& standard deviation of shock in alternative $a$									\\
		$\sigma_{i,j}$ &	& correlation between shocks in alternative $a = i$ and $a=j$ with $i \neq j$ \\

	\end{longtable}
\end{ThreePartTable}

%!TEX root = ../main.tex
%---------------------------------------------------------------------------------------------------
\subsection{Data}\label{Appendix data}\FloatBarrier
%---------------------------------------------------------------------------------------------------
We use the same data as \citetAppndx{Keane.1997}, who derive their sample from the National Longitudinal Survey of Youth 1979 (NLSY79) \citepAppndx{NLSY.2019}. The NLSY79 is a nationally representative sample of young men and women living in the United States in 1979 and born between 1957 and 1964. Individuals were followed from 1979 onwards and repeatedly interviewed about their educational decisions and labor market experiences. Based on this information, individuals are assigned to either working in one of three occupations, attending school, or simply staying at home. The decision period is represented by the school year. The sample is restricted to white men, who turned 16 between 1977 and 1981, and it uses information collected between 1979 and 1987. Thus, the individuals in the sample range in age between 16 and 26 years old.\\

\begin{figure}[ht!]\centering
\scalebox{0.35}{\includegraphics{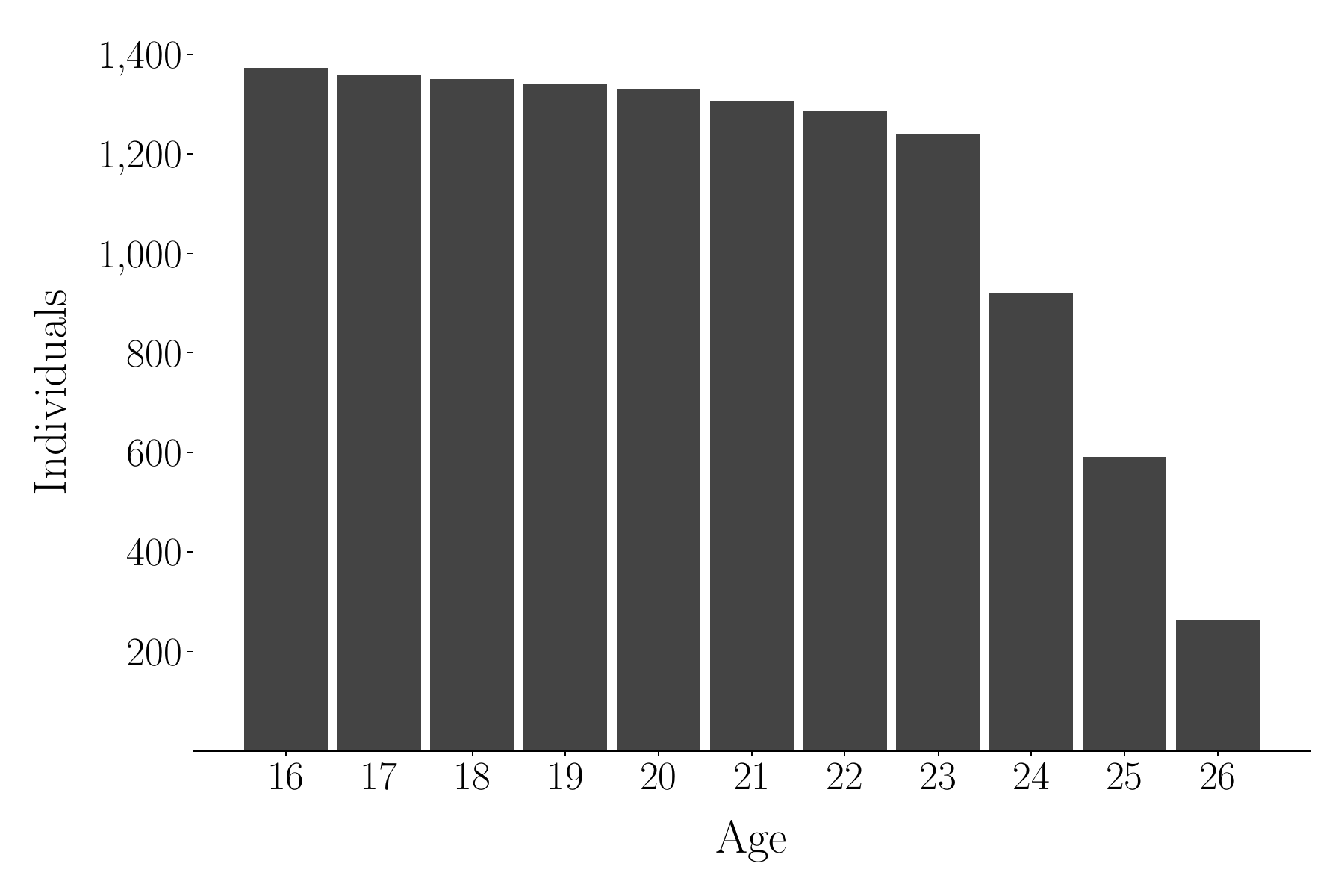}}
\caption{Sample size}\label{Sample size}
\end{figure}

\noindent Figure \ref{Sample size} shows the sample size by age. While the sample initially consists of 1,373 16-year-olds, this value drops to 256, once the sampled individuals reach the age of 26 due to sample attrition, missing data, and the short observation period. Overall, the final sample consists of 12,359 person-period observations.\\

\FloatBarrier

\begin{figure}[ht!]\centering
\scalebox{0.35}{\includegraphics{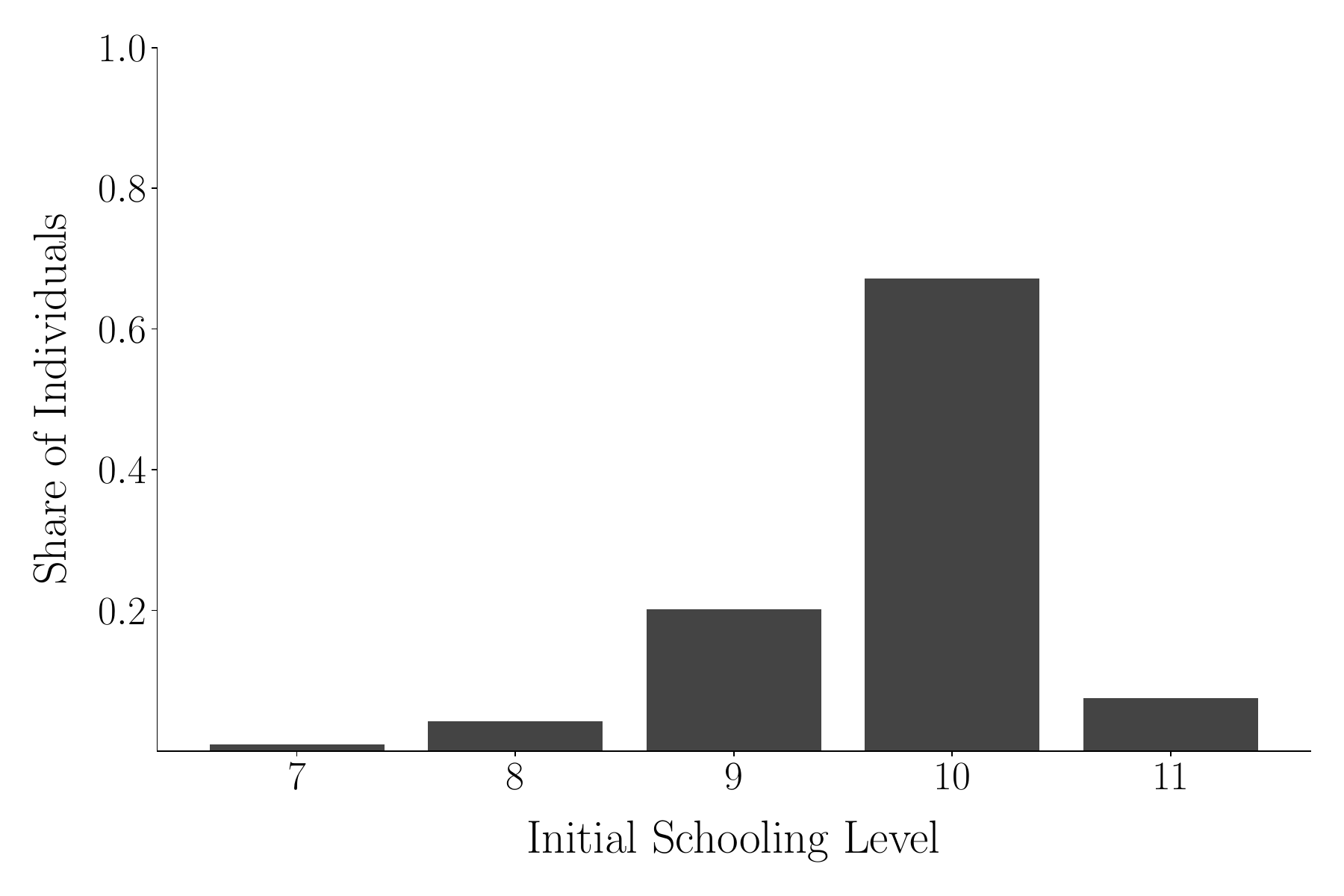}}
\caption{Initial schooling}\label{Initial schooling}
\end{figure}

\noindent Figure \ref{Initial schooling} shows the distribution of initial schooling among individuals at the time they enter the model. The majority of individuals enter the model with ten years of schooling, while about a quarter of the sample has less than ten years of schooling. About $7.5\%$ of individuals already attended school for 11 years.\\

\FloatBarrier

\begin{figure}[ht!]\centering
\scalebox{0.35}{\includegraphics{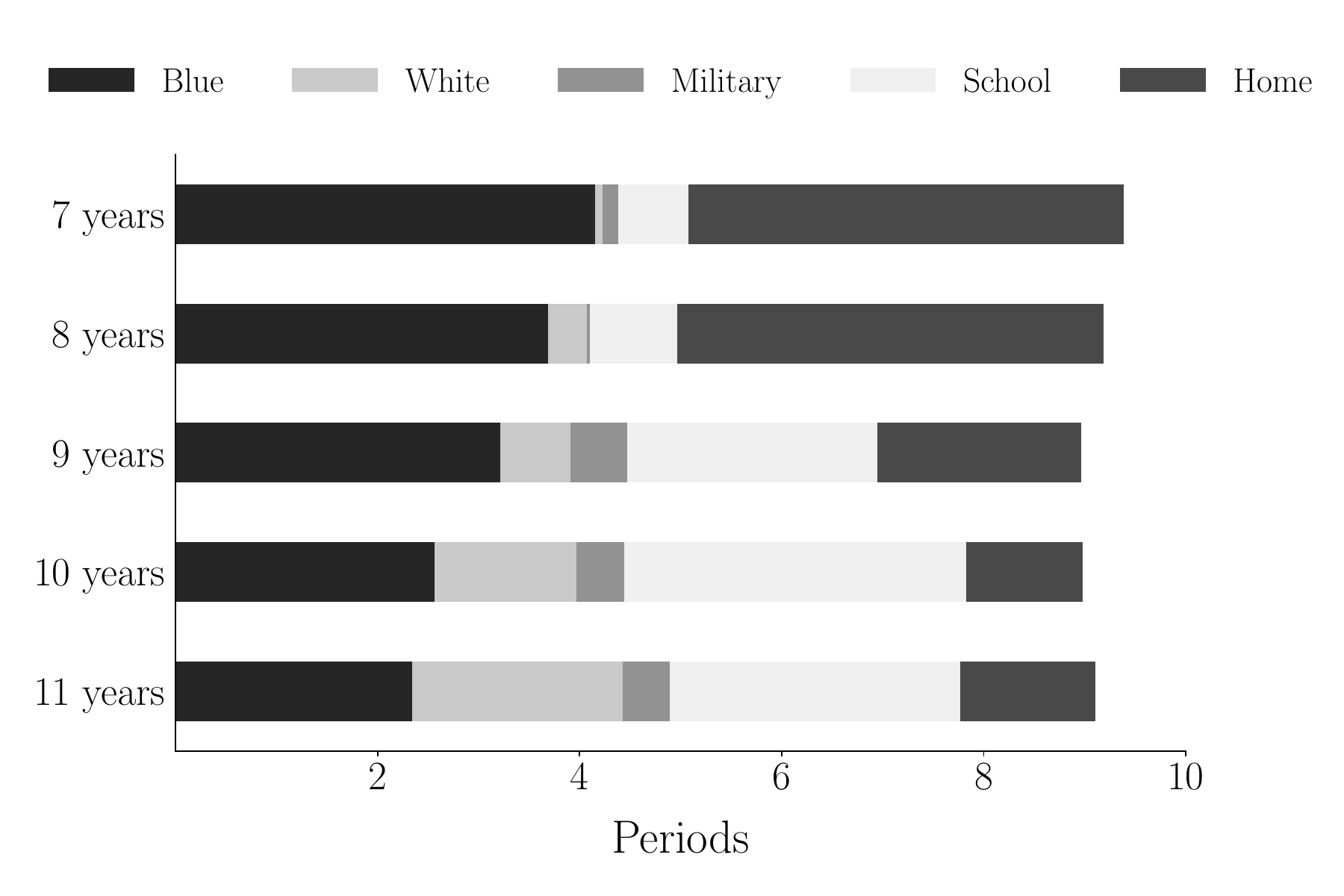}}
\caption{Average choices by initial schooling}\label{Average choices by initial schooling}
\end{figure}

\noindent Figure \ref{Average choices by initial schooling} shows heterogeneity of choices by the level of initial schooling. Individuals who enter the model with only seven years of schooling spend an additional 0.65 years in school after age 16. Consequently, they spend around four years at home. In the event that they are working, it is likely in a blue-collar occupation. When starting with ten years of schooling, then individuals add roughly another three years while in the model. This increase is about half a year more than individuals that start with eleven years.\\

\FloatBarrier

\begin{figure}[h]\centering
\scalebox{0.35}{\includegraphics{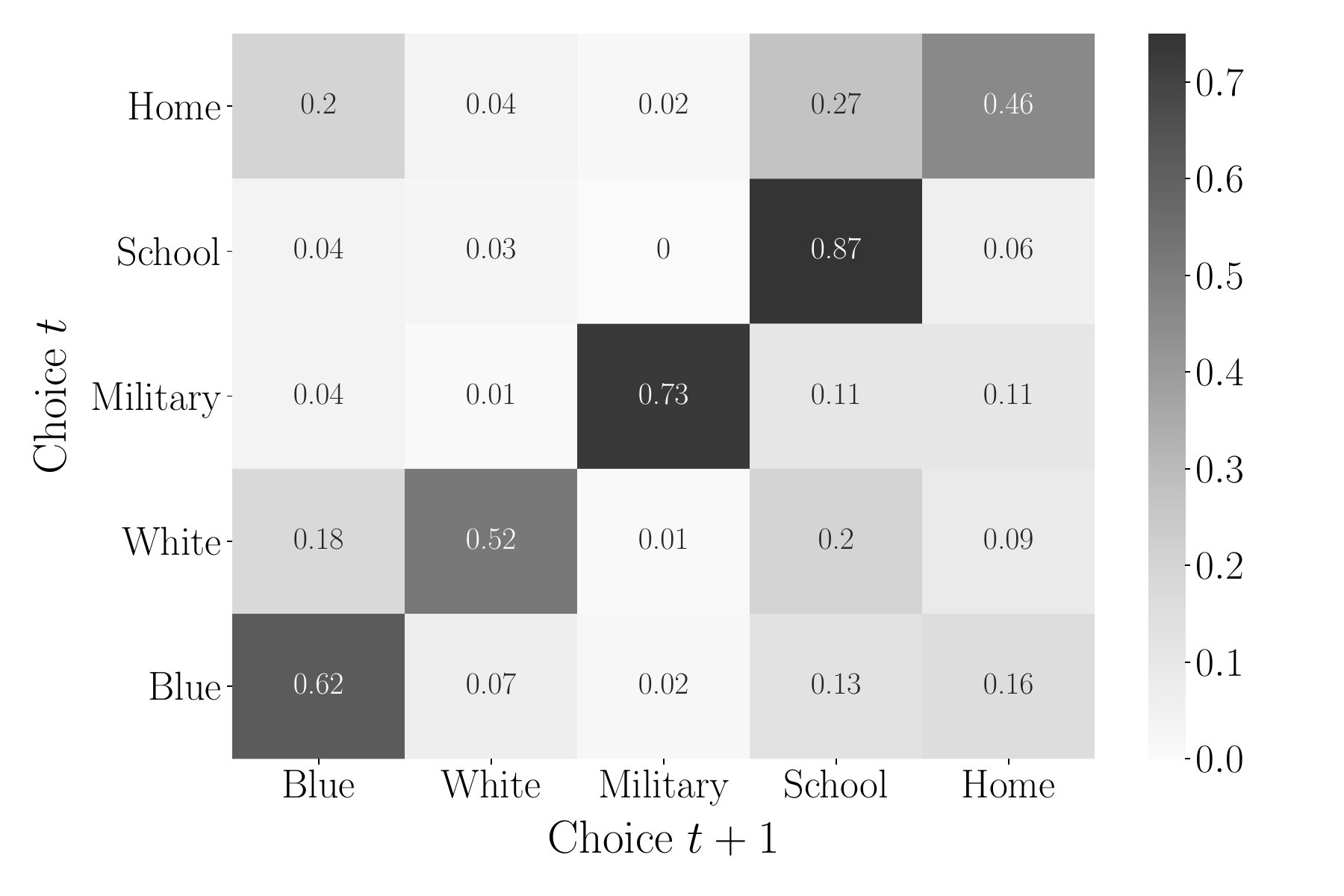}}
\caption{Transition matrix}\label{Transition matrix}
\end{figure}

\noindent Figure \ref{Transition matrix} documents strong persistence in choices over time. For example, among those with a white-collar occupation in $t$, 67\% work in the same occupation in $t + 1$, while $20\%$ switch to a blue-collar job.\\

\FloatBarrier

% !TEX root = ../main.tex
%---------------------------------------------------------------------------------------------------
\subsection{Results}\label{Appendix results}
\begin{figure}[h]\centering
	\subfloat[White-collar]{\scalebox{0.25}{\includegraphics{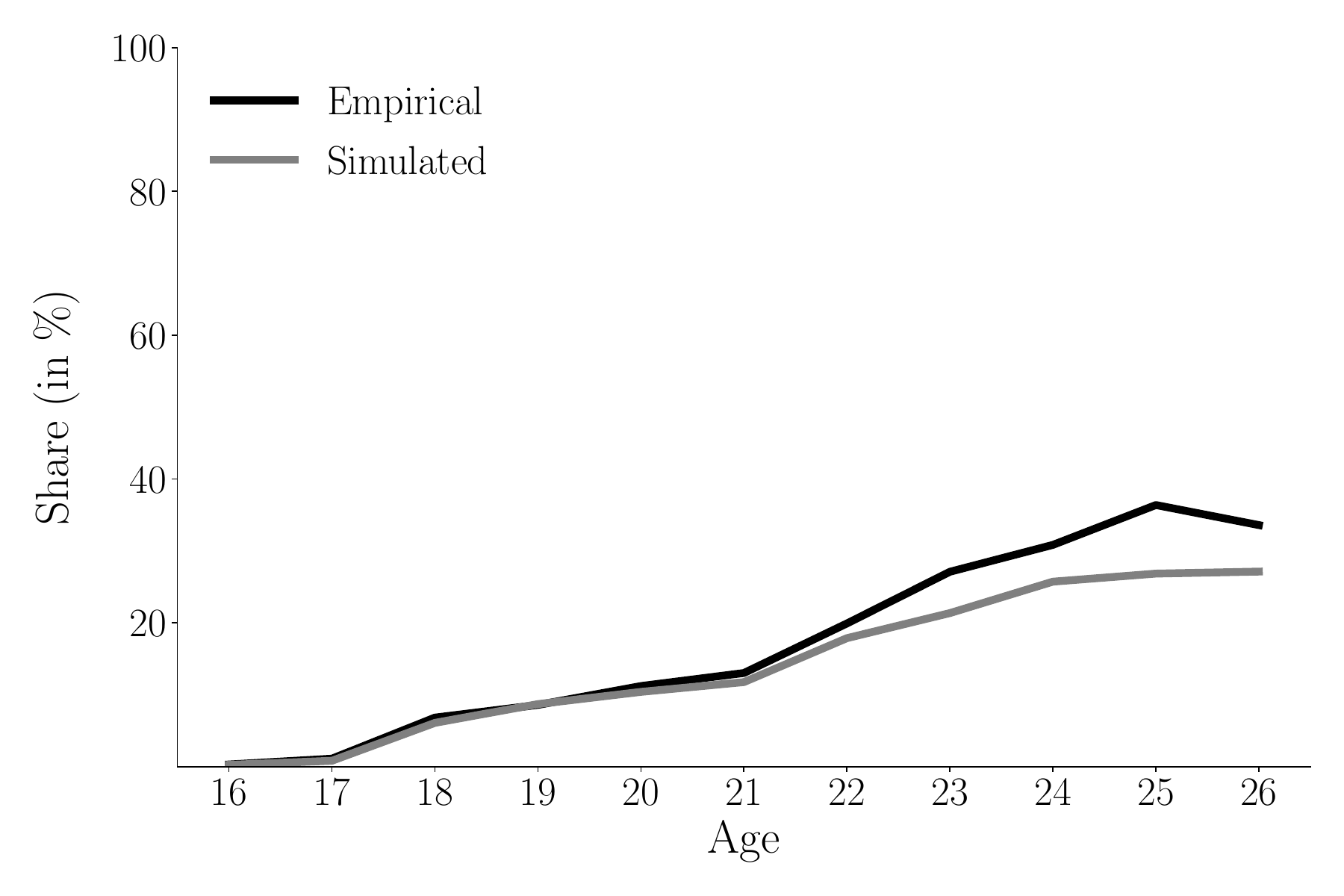}}}
	\subfloat[Military]{\scalebox{0.25}{\includegraphics{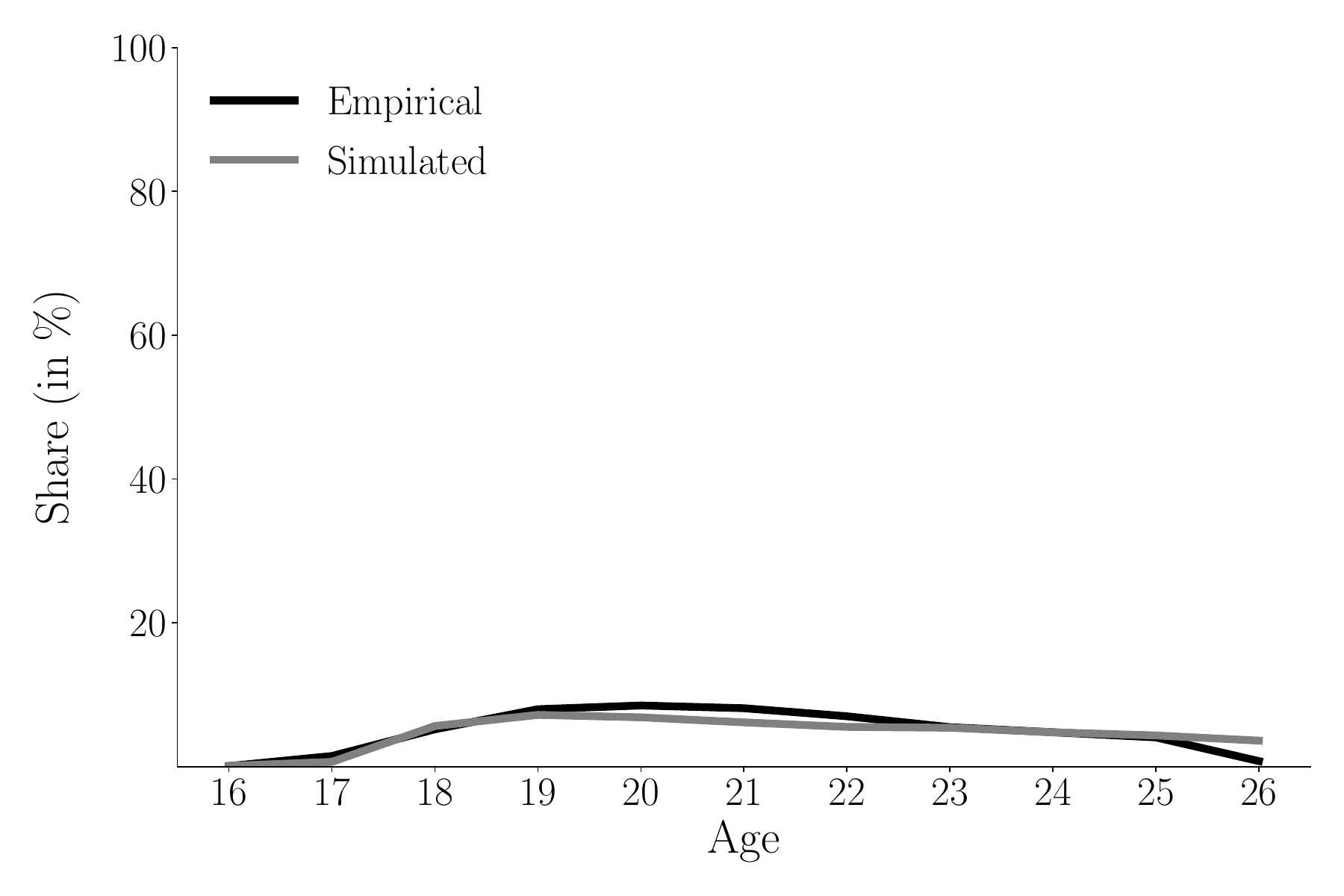}}} \\
	\subfloat[School]{\scalebox{0.25}{\includegraphics{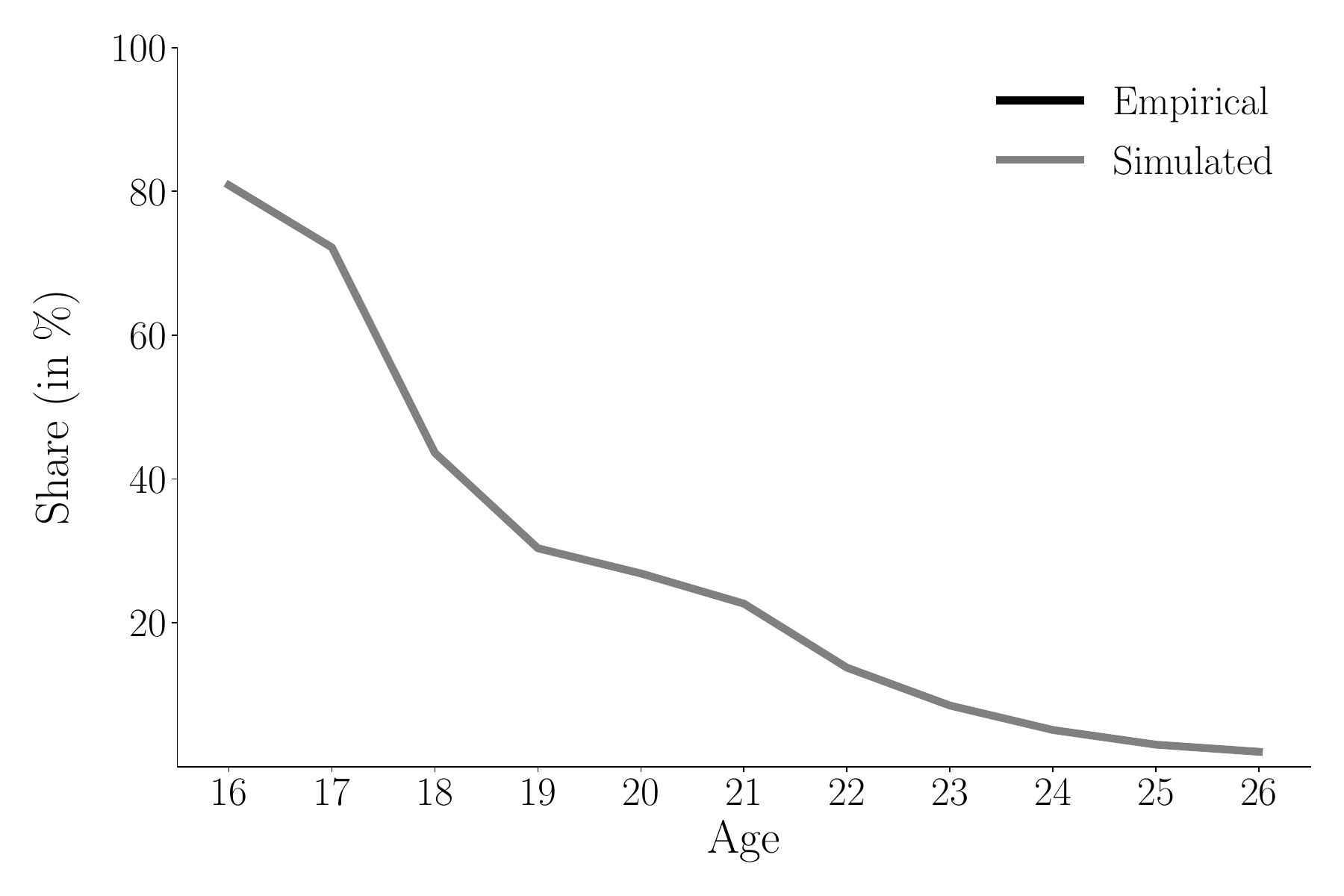}}}
	\subfloat[Home]{\scalebox{0.25}{\includegraphics{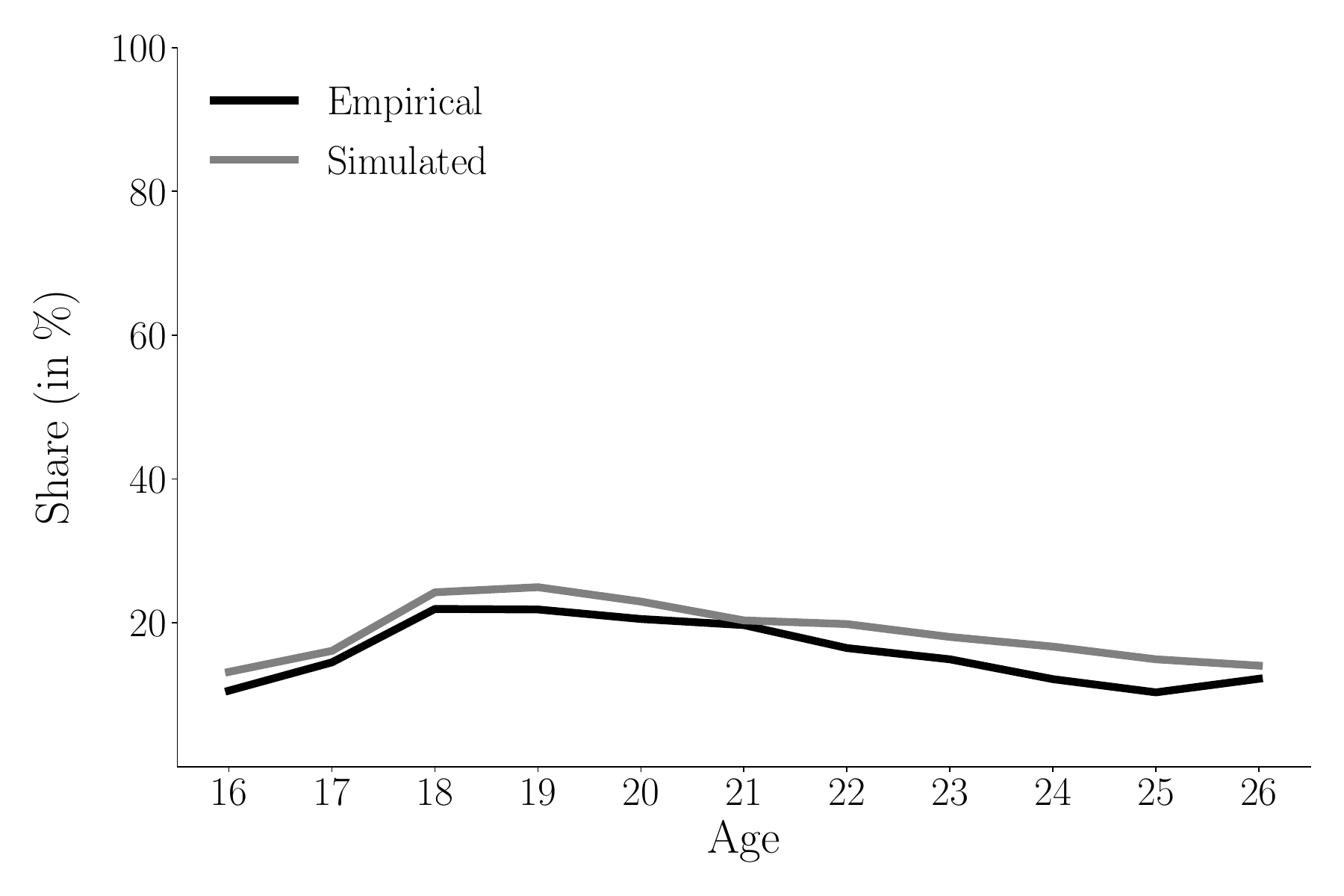}}}
	\caption{Model fit}\label{Model fit appendix}
\end{figure}

\FloatBarrier
%---------------------------------------------------------------------------------------------------
Figure \ref{Model fit appendix} shows further comparisons between the simulated and empirical data. All results from the estimated model are based on $10,000$ individuals.

\FloatBarrier

\begin{figure}[h!]\centering
  \subfloat[Type 1]{\scalebox{0.25}{\includegraphics{fig-policy-average-years-type-0-bw}}}
  \subfloat[Type 2]{\scalebox{0.25}{\includegraphics{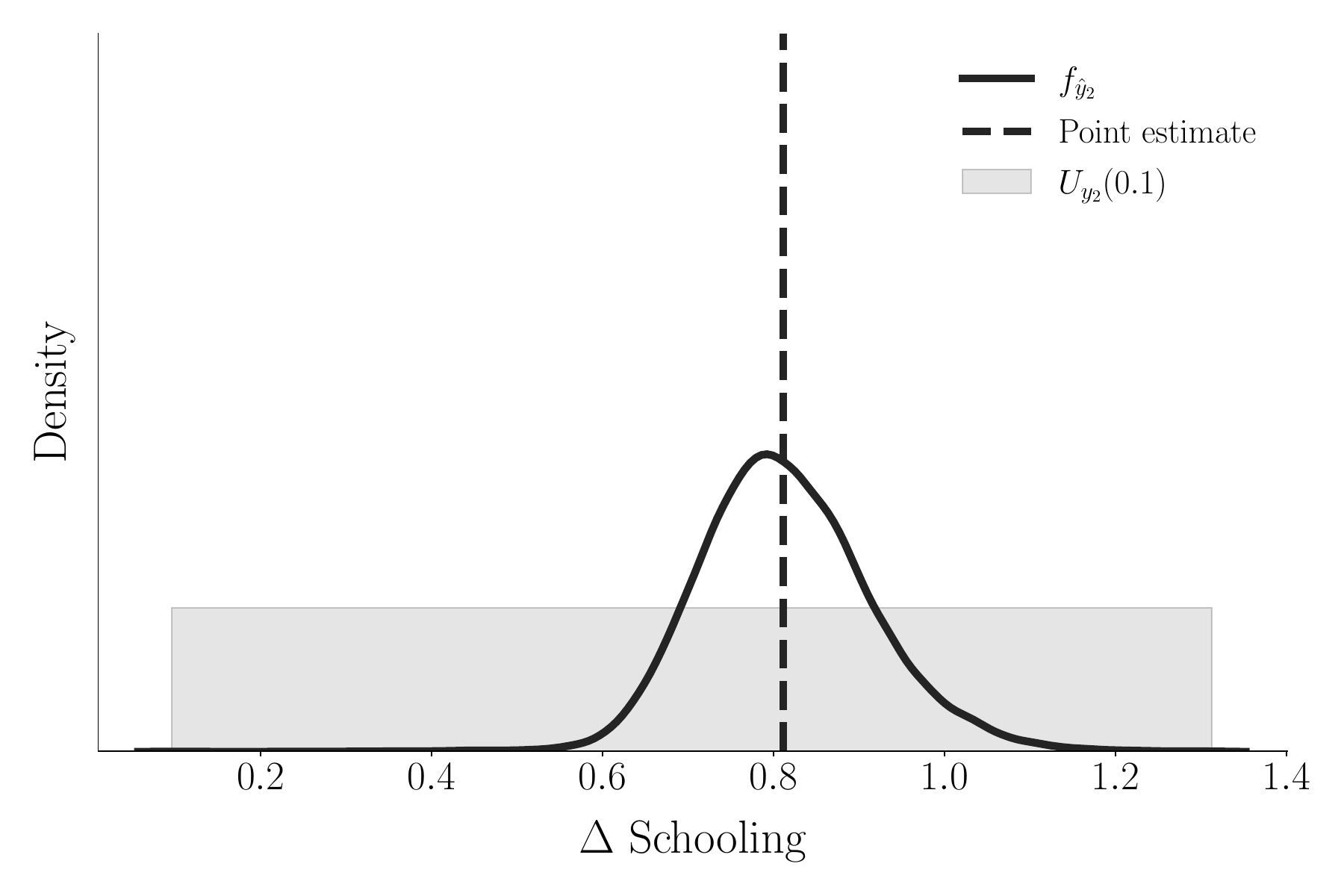}}}\\
  \subfloat[Type 3]{\scalebox{0.25}{\includegraphics{fig-policy-average-years-type-2-bw}}}
  \subfloat[Type 4]{\scalebox{0.25}{\includegraphics{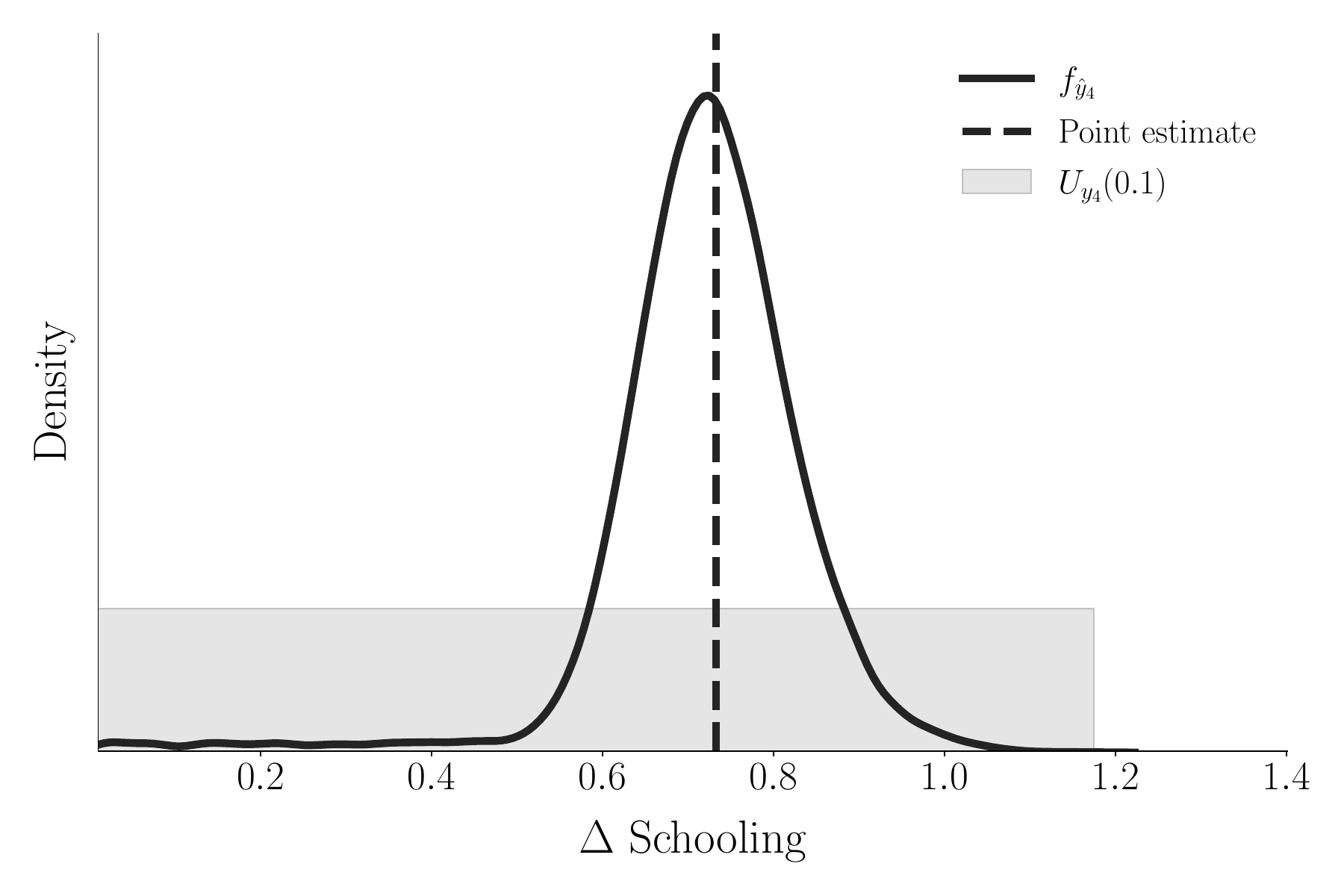}}}
  \caption{Targeted subsidy for all types}\label{Targeted subsidy for all types}
\end{figure}

\noindent Figure \ref{Targeted subsidy for all types} provides the point prediction, its sampling distribution, and the estimated confidence set for the impact of the tuition subsidy on all types. All results are based on $30,000$ draws from the asymptotic normal distribution of our parameter estimates.

\FloatBarrier

\begin{figure}[h!]\centering
\subfloat[Low $\delta$]{\scalebox{0.25}{\includegraphics{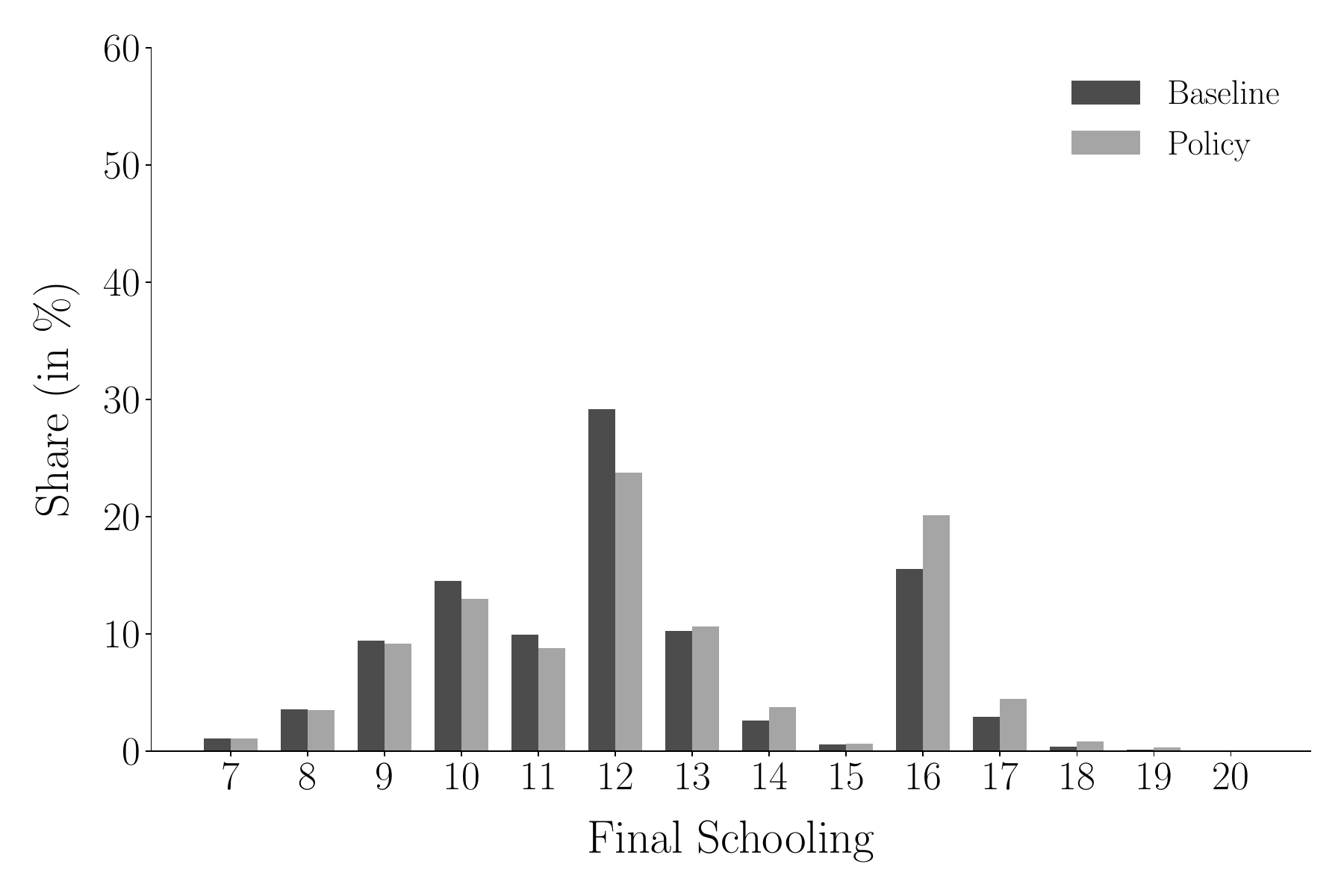}}\label{type 0}}
\subfloat[High $\delta$]{\scalebox{0.25}{\includegraphics{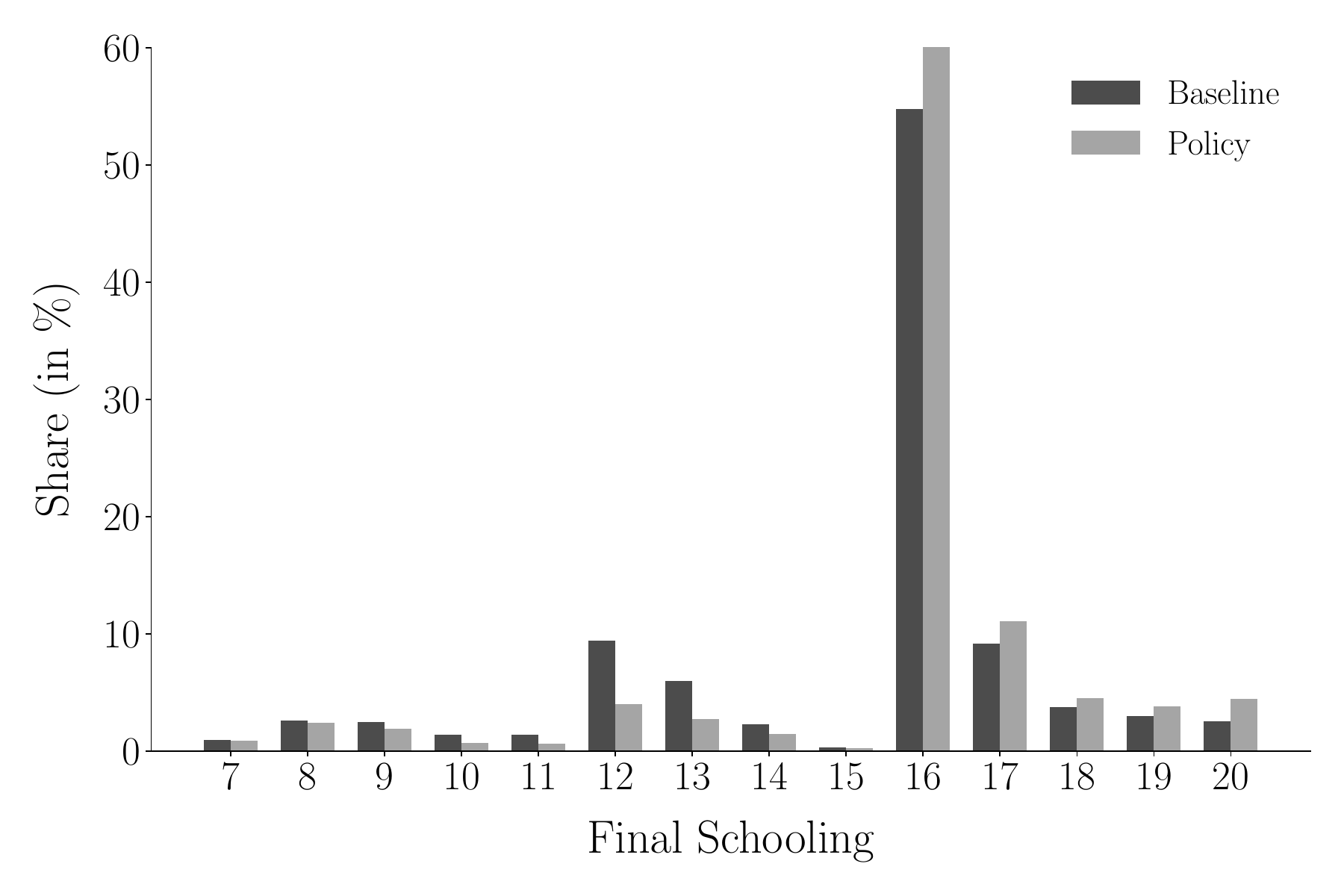}}\label{type 2}}
\caption{Policy impact and time preference}\label{Policy impact and time preference}
\end{figure}

\noindent Figure \ref{Policy impact and time preference} shows the impact of the tuition subsidy at the upper $\delta_H$ and lower $\delta_L$ bound of the estimated confidence set for $\delta$. The results for both scenarios are based on simulated samples of $10,000$ individuals.

\FloatBarrier

%\bibliographystyleAppndx{apalike}
%\bibliographyAppndx{literature}\newpage

\newpage

%---------------------------------------------------------------------------------------------------
%---------------------------------------------------------------------------------------------------
\end{appendices}

\end{document}